\documentclass[useAMS,usenatbib,usegraphicx,twocolumn]{mn2e}
  
\usepackage{float}
\usepackage{aas_macros}
\usepackage{threeparttable}
\usepackage{grffile}
\usepackage{amsmath, amssymb}
\usepackage{multirow}
\usepackage{wallpaper}
\usepackage{mathbbol}
\usepackage{rotating}
\usepackage{blindtext}
\usepackage[T1]{fontenc}
\usepackage{changes}
\usepackage{lscape}

\usepackage{tocloft}
\newcommand{\fat}{\textbf}
\newcommand{\ita}{\textit}

\newcommand{\beq}{\begin{equation}}
\newcommand{\eeq}{\end{equation}}  
\newcommand{\RNum}[1]{\uppercase\expandafter{\romannumeral #1\relax}}

  \title[Magnetised Galaxies]{Global dynamics of the interstellar medium in magnetised disc galaxies}
  
  \author[B.~K\"ortgen \& friends]{Bastian~K\"ortgen$^1$, Robi~Banerjee$^1$, Ralph~E.~Pudritz$^{2,3}$  and Wolfram~Schmidt$^1$\\
  $^{1}$ Hamburger Sternwarte, Universit\"at Hamburg, Gojenbergsweg 112, D-21029 Hamburg, Germany \\
  $^{2}$ Department of Physics and Astronomy, McMaster University, Hamilton, ON L8S 4K1, Canada \\
  $^{3}$ Origins Institute, McMaster University, Hamilton, ON L8S 4K1, Canada  
  }

\date{Released 2019}

\pagerange{\pageref{firstpage}--\pageref{lastpage}} \pubyear{2018}

\begin{document}

\label{firstpage}
\maketitle

\begin{abstract}
Magnetic fields are an elemental part of the interstellar medium in galaxies. However, their impact on gas dynamics and 
star formation in galaxies remains controversial. We use a suite of global magnetohydrodynamical simulations of isolated disc 
galaxies to study the influence of magnetic fields on the diffuse and dense gas in the discs. We find that the 
magnetic field acts in multiple ways. Stronger magnetised discs fragment \ita{earlier} due to the shorter growth time
 of the Parker instability.  Due to the Parker instability in the magnetised discs we also find cold ($T<50\,\mathrm{K}$) and dense 
 \mbox{($n\sim10^3-10^4\,\mathrm{cm}^{-3}$)} gas several hundred pc above/below the 
midplane without any form of stellar feedback. In addition, magnetic fields change the fragmentation pattern. While in the hydrodynamical 
case, the disc breaks up into ring-like structures, magnetised discs show the formation of filamentary entities that extent both 
in the azimuthal and radial direction. These kpc scale filaments become magnetically (super-)critical very quickly and allow for the rapid formation 
of massive giant molecular clouds.  Our simulations suggest that major differences in the behaviour of star formation - due to a varying magnetisation - in galaxies could arise.
\end{abstract}

\section{Introduction}
Magnetic fields are ubiquitous in the interstellar medium (ISM) of galaxies and their role and connection to star formation in galaxies has a long and interesting 
history \citep{Mestel56,Parker66,Mouschovias76,Shu87,Elmegreen89,Padoan99b,Hennebelle08b,Li14,Naab17,KrumholzFederrath19}. The flow of gas between the different phases 
of the interstellar medium is a central question about galactic evolution.  Do magnetic fields affect the formation of molecular clouds?  What regulates the amount of 
gas that exists in the different phases - the warm and cold neutral media (WNM and CNM), the molecular (henceforth CMM-) phase, and the densest star forming gas (SF) within molecular clouds?\\
Observationally, these questions are tackled via measurements of the polarised synchrotron 
radiation of charged particles, polarised thermal (re-)emission of dust grains within molecular clouds or polarised starlight 
(dichroic extinction), via 
the Faraday rotation or the Zeeman effect \citep[e.g.][]{Beck13,Kim16,PattleFissel19}. The former three methods reveal the 
line-of-sight (LOS) averaged magnetic field in the plane of the sky (perpendicular to the LOS), $B_\perp$, while the latter two provide 
information on the (intensity) 
weighted magnetic field along the LOS, $B_\parallel$. As discussed by \citet{Kim16}, Faraday rotation measurements 
suffer from the fact that the obtained rotation measure contains no information about \ita{where} the actual Faraday rotation 
takes place. Hence, information about the spatial contribution to the average $B_\parallel$ is missing. Furthermore, 
different regions can show a varying degree of Faraday rotation. As a consequence, the obtained signal might have become 
depolarised, leading to an underestimate of $B_\parallel$ of the region of interest. A similar trend is seen in polarisation studies due to strongly 
tangled fields along the LOS or inefficient dust alignment \citep{PattleFissel19}. However, despite the fact that the observational technique used mostly depends on the region of interest,  
results from these different methods agree on the fact that the magnetic field seems to play a key role for the dynamics within galaxies. \\
Measurements of the Zeeman effect allow for the only way to directly 
determine the magnetic field \ita{strength} \citep[or more precisely its LOS component,][]{Heiles05,Crutcher10}. H\,I absorption line surveys by \citet{Heiles05} 
reveal that the median LOS field amplitude is 
$B_\mathrm{median}\sim(6\pm1.8)\,\mu\mathrm{G}$. As the authors point out, these data can be best ascribed to the CNM (with $T\lesssim200\,\mathrm{K}$). The authors 
further find that the magnitude $B_\mathrm{median}\sim\mathrm{const}$, which hints towards gas motion along magnetic field lines \citep[see e.g. review by][]{Crutcher12,Hennebelle12}. This has recently been supported by analyses of data from  
the LAB \citep{Kalberla05} and GASS \citep{McClureGriffiths09} H\,I surveys, which find diffuse H\,I filaments in the Galaxy that are well aligned with the magnetic field, where the latter is inferred from starlight polarisation \citep{Clark14,Clark15,Kalberla16}.\\
As pointed out by \citet{Kalberla16}, the typical volume density of their observed H\,I filaments is of the order $n\gtrsim10\,\mathrm{cm}^{-3}$. In the denser gas, magnetic field strengths can be 
inferred by the Zeeman effect of OH \citep[with $n_\mathrm{typical}\sim10^{2-4}\,\mathrm{cm}^{-3}$,][]{Troland08} or CN \citep[$n_\mathrm{typical}\sim10^{5-6}\,\mathrm{cm}^{-3}$,][]{Falgarone08}. In such 
regions, the field strengths are typically $B_\mathrm{LOS}\sim50-1000\,\mu\mathrm{G}$.\\
\citet[][see also review by \citet{Crutcher12}]{Crutcher10} studied the magnetic field - density relation as obtained from Zeeman measurements. The authors showed that the large compendium of 
observations indicates that the field strength stays approximately constant up to $n\sim300\,\mathrm{cm}^{-3}$. Above this threshold, the field magnitude increases as 
$B\propto n^{\alpha}$, where $\alpha\sim1/2-2/3$, due to compression of the field via (almost spherical) gravitational 
contraction. Comparing magnetic and gravitational energies in the different regimes reveals that the diffuse (H\,I) gas is magnetically dominated (i.e. sub-critical with $E_\mathrm{grav}/E_\mathrm{mag}<1$), 
while the dense (from OH and CN) gas appears to be super-critical ($E_\mathrm{grav}/E_\mathrm{mag}>1$) by factors of $\sim2-3$. Since the diffuse ISM is magnetically dominated, $\mu\mathrm{G}$-field strengths must be generated by a
non-adiabatic (that is, by a non-compressive) process, such as the turbulent (small-scale) dynamo \citep{Federrath11a,Schober12a,Pakmor13,Steinwandel19} or an
$\alpha\Omega$-dynamo, which acts on larger scales due to (global) galactic 
shear \citep[see e.g.][and references therein]{Beck13,Klein15,Beck16,Steinwandel19}. As noted in 
\citet[][see also \cite{Kotarba10}]{Kotarba09}, the 
$\alpha\Omega$-dynamo is capable of amplifying a seed magnetic field to the $\mu\mathrm{G}$-level within only a 
few disc rotations. This agrees with more recent analytical and numerical findings by 
\citet[][though these authors 
focus on the study of the galactic halo magnetic field]{BeckAM12}, who further 
highlight that the turbulence, relevant for the dynamo action, can be driven either by supernova feedback or 
even by gravitational collapse of large scale structures. \\
As reviewed in \citet[][see also \citet{Beck13,Beck16}]{Beck15}, magnetic pressure usually dominates over thermal 
pressure on large, galactic scales with upper limits of $P_\mathrm{th}/P_\mathrm{mag}\equiv\beta<1$. Observations of dense filaments and cores in the 
Milky Way reveal a similar trend with $\beta\ll1$ \citep{Busquet16,Santos16}, despite the large observational biases due to 
determining the volume density from column densities. As further discussed in \citet{Soler13} and, more recently, 
by the \citet{Planck16c}, the observed orientation of the magnetic field (from dust polarisation) with respect to gradients in column 
density indicates a dynamically significant field. \citet{Soler17b} point out that the observed patterns can only be explained 
by regimes with $\beta<1$ and sub- to trans-alfv\'{e}nic turbulence. These latter conditions also seem to hold for magnetic fields in external galaxies, as revealed 
through comparison of the field in the diffuse gas with the one within molecular clouds \citep[e.g. in M33,][]{Li11}.\\
Magnetic fields are believed to affect the star formation properties of the dense, molecular gas. However, there is currently only a limited repertoire of observations that 
relate the star formation properties to the magnetic field. Recently, \citet{Li17} reported that the star formation rate per unit cloud mass of nearby molecular clouds decreases 
with increasing angle between the magnetic field and the cloud's major axis. However, as argued in \citet{KrumholzFederrath19}, this might have its roots in the 
relation of the field and the star formation process with density. \citet{Tabatabaei18} studied the effect of the magnetic field on the star formation process in the center of NGC\,1097. 
They found that the dimensionless star formation rate per free-fall time decreases (by a factor of $\sim3$) with increasing magnetic field strength ($B$ increased by a factor of $1.25$).\\
Although there is profound evidence in numerical simulations that magnetic fields significantly affect the star formation 
properties of molecular clouds \citep[e.g.][]{Vazquez11a,Koertgen15,Federrath15b}, its long-term influence on the gas dynamics on larger scales is less certain \citep{Khoperskov18}. 
For example, \citet{Girichidis18} report a quite significant loss of magnetic flux via galactic winds and outflows driven by 
supernova explosions over a period of several tens of Myr. The net effect of the magnetic field was shown to be 
negligible and the authors concluded that the major impact of the field is in delaying gas fragmentation. 
\citet{Koertgen18L} 
performed global galactic scale simulations of disc galaxies and showed that magnetised galaxies fragment due to the 
Parker instability. The resulting fragments, which formed in the magnetic valleys via convergence of gas flows along 
magnetic field lines, were observed to be magnetically supercritical. These gas motions parallel to the magnetic field were  
several kpc long, i.e. comparable in size to the characteristic wavelength of the Parker instability, $\sim4\pi H$ with $H$ being the scale height. 
This accumulation scenario, already proposed by \citet{Mestel56}, in which mass is being collected in magnetic valleys from large distances, 
certainly provides the conditions for the formation of even the most massive molecular clouds (so-called GMCs) out of a strongly magnetised interstellar medium.  \\
Given that the magnetic field dominates the energetics of the WNM, but is sub-dominant in the physics of molecular gas, 
magnetohydrodynamic effects can play an important roll in controlling the global evolution of the ISM.
In this paper, we therefore present our findings from a study of galactic scale simulations with 
a maximum spatial resolution of about 20\,pc. To focus on the impact of the magnetic field on the distribution and evolution of the various gas phases,
 we do not include feedback effects from star formation. In contrast to previous studies of magnetic fields in galaxies 
 (see references given above), we 
start with a saturated magnetic field, as we are more interested in its late-time effects. Section \ref{secIC} introduces the 
initial conditions and discusses the numerical details. In section \ref{secRes} we present our results and we conclude this 
paper in section \ref{secSum}.

\section{Initial conditions and numerics}
\label{secIC}
\subsection{Initial conditions}
We initialise the disc in the centre of a cubic domain with edge length of \mbox{$L_\mathrm{box}=40\,\mathrm{kpc}$}. Following \citet[][see also \cite{Koertgen18L}]{Tasker09}, the density profile of a thin disc is used:
\beq
\varrho(R,z)=\frac{\kappa c_\mathrm{s}\sqrt{1+\frac{2}{\beta}}}{2\pi G Q_\mathrm{eff}H(R)}\mathrm{sech}^2\left(\frac{z}{H(R)}\right),
\eeq
with $\kappa,c_\mathrm{s}$ and $\beta=2c_\mathrm{s}^2/v_\mathrm{a}^2$ being the epicyclic frequency, the isothermal sound speed and the plasma-$\beta$, respectively. Furthermore, 
$Q_\mathrm{eff}=\kappa\left(c_\mathrm{s}^2+v_\mathrm{a}^2\right)^{1/2}\big/\pi G\Sigma$ is the 
effective Toomre-parameter (accounting for thermal and magnetic support) and \mbox{$H(R)=R_\odot(0.0085+0.01719R/R_\odot+0.00564(R/R_\odot)^2)$} is the radially increasing scale height of the disc and 
we take \mbox{$R_\odot=8.5\,\mathrm{kpc}$}. The disc initially extends out to \mbox{$R_\mathrm{out}=10\,\mathrm{kpc}$} and the 
density is constant and negligibly small for \mbox{$R>R_\mathrm{out}$} and \mbox{$|z|>H(R)$}. The resultant 
disc mass is of the order of $M_\mathrm{disc}\sim10^{10}\,\mathrm{M}_\odot$, i.e. comparable to the mass 
of the LMC.\\
We strive to study the impact of magnetic fields on the evolution of disc galaxies. Hence, we incorporate a magnetic field into the disc. For the sake of simplicity, we only use the $m=0$ mode, which gives 
an entirely toroidal field. In conjunction with recent Zeeman observations \citep[see e.g.][]{Crutcher12} we set the dependence of the magnetic field on the gas density as $B\propto n^\alpha$ 
with $\alpha=0.5$. This form also has the advantage that the plasma-$\beta$ of the disc is initially constant. We here point out that this scaling relation is also used for gas densities below 
\mbox{$n\sim300\,\mathrm{cm}^{-3}$}, whereas observations indicate a rather constant field strength in this regime \citep{Crutcher10,Crutcher12,Beck15}. The magnetic field surrounding the galaxy is vanishingly small.\\
Disc fragmentation due to gravitational instability occurs when the Toomre-parameter, $Q$, drops below unity. In order to avoid rapid fragmentation, we set the initial Toomre-parameter to $Q_\mathrm{eff}=2$ for $0.5<R/\mathrm{kpc}<8.5$ and 
$Q_\mathrm{eff}=20$ elsewhere. This ensures that the disc has time to evolve dynamically and that the innermost part near the centre is stable for a long time, as this region is less well resolved. The 
transition to $Q_\mathrm{eff}<1$ can be achieved by gas cooling, which is based on the prescription by \citet[][with corrections by \cite{Vazquez07}]{Koyama02}. The fitted functions for optically thin 
cooling and heating read
\beq
\begin{split}
\frac{\Lambda(T)}{\Gamma} &=10^7\mathrm{exp}\left(\frac{-1.184\times10^5}{T+1000}\right)\\
&\quad +1.4\times10^{-2}\sqrt{T}\mathrm{exp}\left(\frac{-92}{T}\right)\,\mathrm{cm}^{3},
\end{split}
\eeq
where $\Lambda(T)$ is the temperature-dependent cooling rate and $T$ the temperature in Kelvin, and
\beq 
\Gamma = 2\times10^{-26}\,\mathrm{erg\,s}^{-1},
\eeq
with the (constant) heating rate $\Gamma$. The latter includes heating from cosmic and soft X-rays, by the photoelectric effect as well as the formation and dissociation of H$_2$ \citep{Wolfire95,Koyama00,Vazquez07}.\\
In addition to the self-gravitational potential of the gas, we use a fixed, stationary external logarithmic potential of the form
\beq
\Phi_\mathrm{ext}=\frac{1}{2}v_0^2\mathrm{ln}\left\{\frac{1}{R_\mathrm{c}^2}\left[R_\mathrm{c}^2+R^2+\left(\frac{z}{q}\right)^2\right]\right\},
\eeq
which accounts for old stars and dark matter. Here, \mbox{$R_\mathrm{c}=0.5\,\mathrm{kpc}$} is the core radius, \mbox{$q=0.7$} the axial ratio and \mbox{$v_0=200\,\mathrm{km/s}$}. This potential gives a flat rotation curve 
\beq
v_\mathrm{rot}=v_0\frac{R}{\sqrt{R_\mathrm{c}^2+R^2}}.
\eeq
We emphasise here that this static potential does not capture the full dynamics of the system due to the missing 
gravitational back-reaction of the baryons onto the dark matter component, such as flattening of the dark 
matter profile near the center of the galaxy. More complex systems, including e.g. a spiral potential which will produce 
large scale spiral arms \citep[see e.g.][]{Dobbs06} or a separately treated stellar component, will be analysed in a future study. An overview of the initial conditions is given in \mbox{Table \ref{tabIC}}.
\begin{table}
\centering
 \caption{List of performed simulations.}
  \begin{tabular}{l|cc}
    \hline
    \hline
    \fat{Run name}	&\fat{plasma-$\beta$}	&$B(R=8\,\mathrm{kpc})$	\\ 
			&			&$[\mu\mathrm{G}]$	\\ 
    \hline
    Hydro		&$\infty$		&0		\\
    Beta10		&10			&2		\\
    Beta5		&5			&3		\\
    Beta1		&1			&10		\\
    Beta0.5		&0.5			&14		\\
    Beta0.25	&0.25		&23		\\
    \hline
    \hline
  \end{tabular}
 \label{tabIC}
\end{table}

\subsection{Numerics}
For our study of galaxy evolution we use the \textsc{flash} code \citep[v4.2.2,][]{Dubey08}. The ideal 
(that is, without any diffusion term added) magnetohydrodynamic (MHD) equations are solved every
 timestep using a HLL5R Riemann solver 
\citep[][]{Waagan11}. Please note that we do not change the numerical solver, when running a pure hydro 
simulation, but simply set the initial magnetic field to zero. To ensure the \mbox{$\nabla\cdot$\fat{B}=0} constraint, we use a hyperbolic cleaning scheme 
\citep[based on][see \cite{Waagan11} for further details.]{Dedner02}. Poisson's equation for the self-gravity of the gas is solved with a Barnes-Hut tree solver \citep[optimised for GPU,][]{Lukat16}. The root grid has a resolution of 
\mbox{$\Delta x_\mathrm{root}=625\,\mathrm{pc}$} and we allow for additional five levels of refinement, using the adaptive mesh refinement technique \citep[AMR,][]{Berger84}. This gives a maximum resolution of 
\mbox{$\Delta x_\mathrm{max}=19.5\,\mathrm{pc}$}. The numerical mesh is refined when the local Jeans length is resolved with less than 32 grid cells and it is de-refined when more than 64 cells 
resolve the Jeans length. Due to this procedure, almost the entire disc is refined to the highest level of refinement at $t=0$. 
 The advantage of the AMR technique becomes significant once the discs have 
started 
to fragment. In order to avoid artificial fragmentation on the highest level of refinement due to violation of the Truelove-criterion \citep{Truelove97}, we introduce an artificial pressure 
term on the highest level of refinement, which is adjusted so that the local Jeans length is resolved with at least four grid cells. 
We like to point out that the high initial resolution ensures that the gas scale height as defined above is resolved with 
10-15 grid cells from $R\sim5\,\mathrm{kpc}$ on. In conjunction with the Truelove-criterion and the artificial pressure 
term, fragmentation of the disc is resolved in the major part of the disc. We use outflow boundary conditions for the magnetohydrodynamics, which 
allows gas to enter or leave the numerical domain, and isolated boundaries for the gravity. This choice of boundary conditions for the MHD implies that angular momentum will not be conserved due to matter 
being allowed to enter or leave the computational domain. However, we observe an angular momentum variation of 
$\lesssim20\,\%$ at late times for both MHD and hydro simulations. Before disc fragmentation, these variations are 
$<1\,\%$.

\section{Results}
\label{secRes}
Before going into a detailled discussion of the time evolution of the galaxies, we define two time scales. These are 
\begin{itemize}
 \item $t_\mathrm{frag}$\\[0.1cm]
  The time of disc fragmentation into individual objects. In addition to a coarse identification by eye, we check whether individual clouds are found by a simple clump-finding algorithm. This algorithm 
  identifies spatially connected objects based on a minimum threshold density of $n=200\,\mathrm{cm}^{-3}$.  We provide an overview of the times of fragmentation for all discs in 
  Table~\ref{tabFrag}.\\
  
  \item $t_\mathrm{onerot}=t_\mathrm{frag}+\frac{2\pi}{\Omega(R=8\,\mathrm{kpc})}$\\[0.2cm]
  The disc has completed a full orbit at $R=8\,\mathrm{kpc}$, starting at $t=t_\mathrm{frag}$. A full orbit at this galactocentric distance takes about $t\sim220\,\mathrm{Myr}$. We emphasise that, 
  during this period, the disc will be drastically influenced by stellar feedback, which we do not take into account in this study.
\end{itemize}

\begin{table}
	\centering
		\caption{Times at which the discs fragment first into individual clouds (in order of increasing $t_\mathrm{frag}$). First column denotes the initial ratio of thermal to 
		magnetic pressure. The second and third column show the fragmentation time in absolute units and normalised to the orbital time at $R=8\,\mathrm{kpc}$ 
		(\mbox{$T_\mathrm{orbit,R=8\,kpc}\sim220\,\mathrm{Myr}$}), respectively.}
		\begin{tabular}{ccc}
			\hline
			\hline
			\fat{plasma}-$\beta$	&$t_\mathrm{frag}$	&$t_\mathrm{frag}/T_\mathrm{orbit,R=8\,kpc}$\\
											&$\left[\mathrm{Myr}\right]$	&\\
			\hline
			$\infty$				&146		&0.65\\
			0.25					&240		&1.07\\
			0.5					&300		&1.34\\
			5					&320		&1.43\\
			10					&330		&1.47\\
			1					&424		&1.90\\
			\hline
			\hline
		\end{tabular}
		\label{tabFrag}
\end{table}

\subsection{Disc stability}
\begin{figure*}
 \centering
 \includegraphics[width=0.45\textwidth,angle=-90]{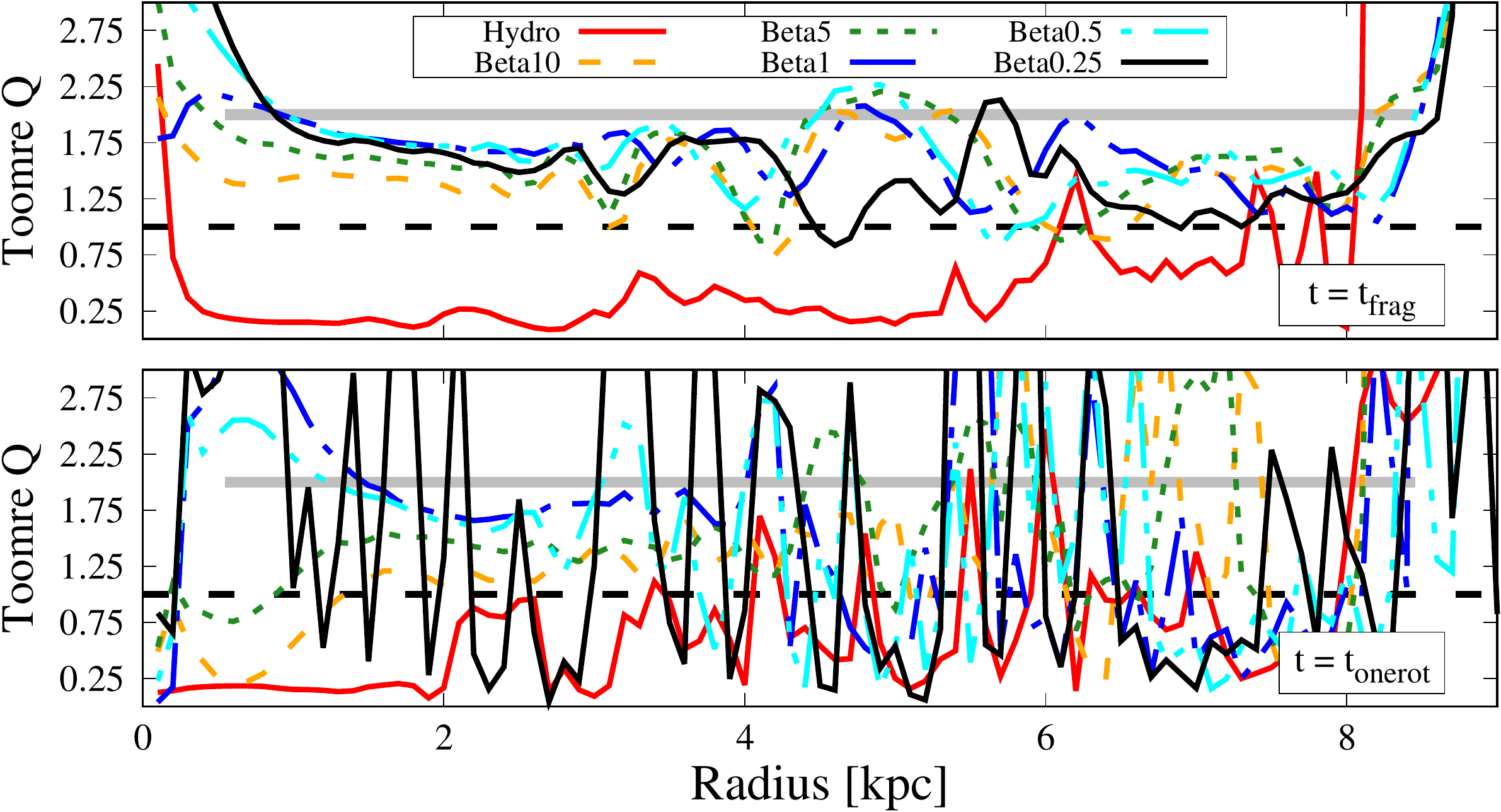}
 \caption{Radial profiles of the density-weighted Toomre parameter at times $t_\mathrm{frag}$ (top) and $t_\mathrm{onerot}$ (bottom). The MHD discs all show $Q\gtrsim1$, but yet they are fragmenting. This is indicative of 
 the Parker instability as the relevant mode of fragmentation in these discs. For comparison, the hydrodynamic disc shows $Q<1$. Already after one addition rotation, $Q<1$ for all discs, with strong 
 fluctuations across the disc. The solid grey line in the back denotes the initial Toomre parameter, $Q=2$, of the main disc. The dashed black lines highlights the critical Toomre parameter, $Q=1$.}
 \label{figToomre}
\end{figure*}

Fig.~\ref{figToomre} shows radial profiles of the Toomre parameter of all discs at times $t_\mathrm{frag}$ and $t_\mathrm{onerot}$. 
In both sub-plots, the grey solid line denotes the initial $Q$-value ($Q=2$) of the main disc. The striking difference between the hydrodynamic and the magnetised discs is the value of the Toomre parameter. 
Usually, a value $Q<1$ indicates that discs are susceptible to gravitational instability. This is the case for the hydrodynamic control disc, but not for the MHD discs. The latter discs show Toomre 
values $Q\sim1-2$, where the upper limit is exactly the initial value. But the definition of $t_\mathrm{frag}$ is such that the disc is observed to undergo fragmentation. Hence, from this figure it is 
clear that the magnetised discs do not fragment according to a gravitational instability. As emphasised in \citet{Koertgen18L}, the magnetised discs fragment due to the Parker instability (PI). The 
overdensities in the magnetic valleys are, at least to some extent, indicated by the drops in the Toomre $Q$, but the overall rather periodic changes of magnetic valleys and hills keeps the average 
Toomre parameter per annulus approximately constant.\\
At time $t_\mathrm{onerot}$, all discs show a highly fluctuating radial profile of the stability parameter. Regions of enhanced stability co-exist with annuli, where the gas is observed to be highly 
unstable. These latter regions are primarily individual molecular clouds that have formed via a warm-cold phase transition in the fragmenting disc. 
A closer view reveals that the inner parts up to $R\sim4\,\mathrm{kpc}$ of the magnetised discs with $\beta\leq 1$ still possess $Q\sim2$. This indicates that these discs have not yet 
fragmented in their inner regions.
\subsection{Disc morphology and fragmentation properties}
\begin{figure*}
  \begin{tabular}{ccc}
  \fat{\Large{Hydro}}	&\fat{\Large{Beta1}}	&\fat{\Large{Beta0.25}}\\
   \includegraphics[width=0.33\textwidth]{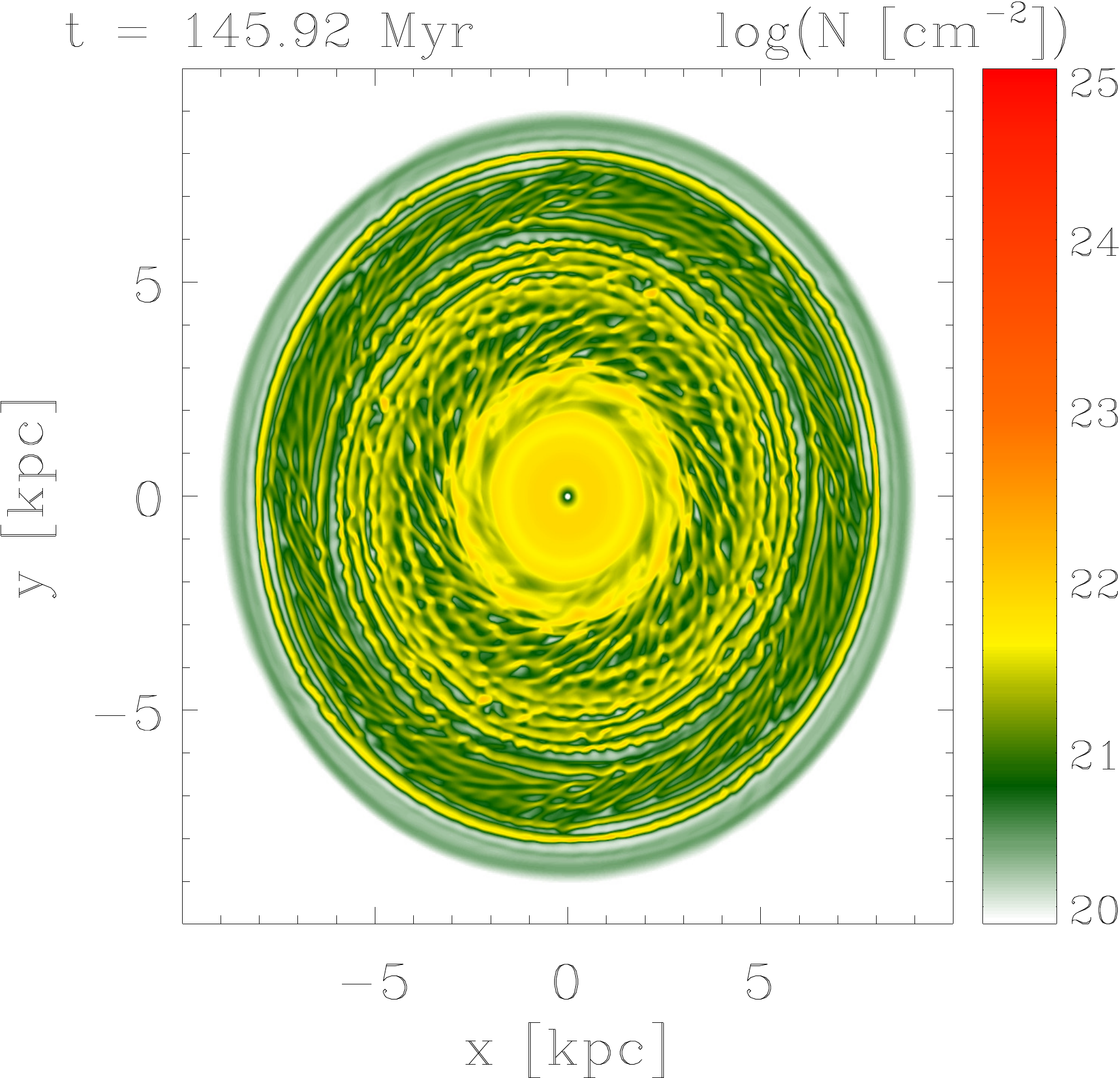}&\includegraphics[width=0.33\textwidth]{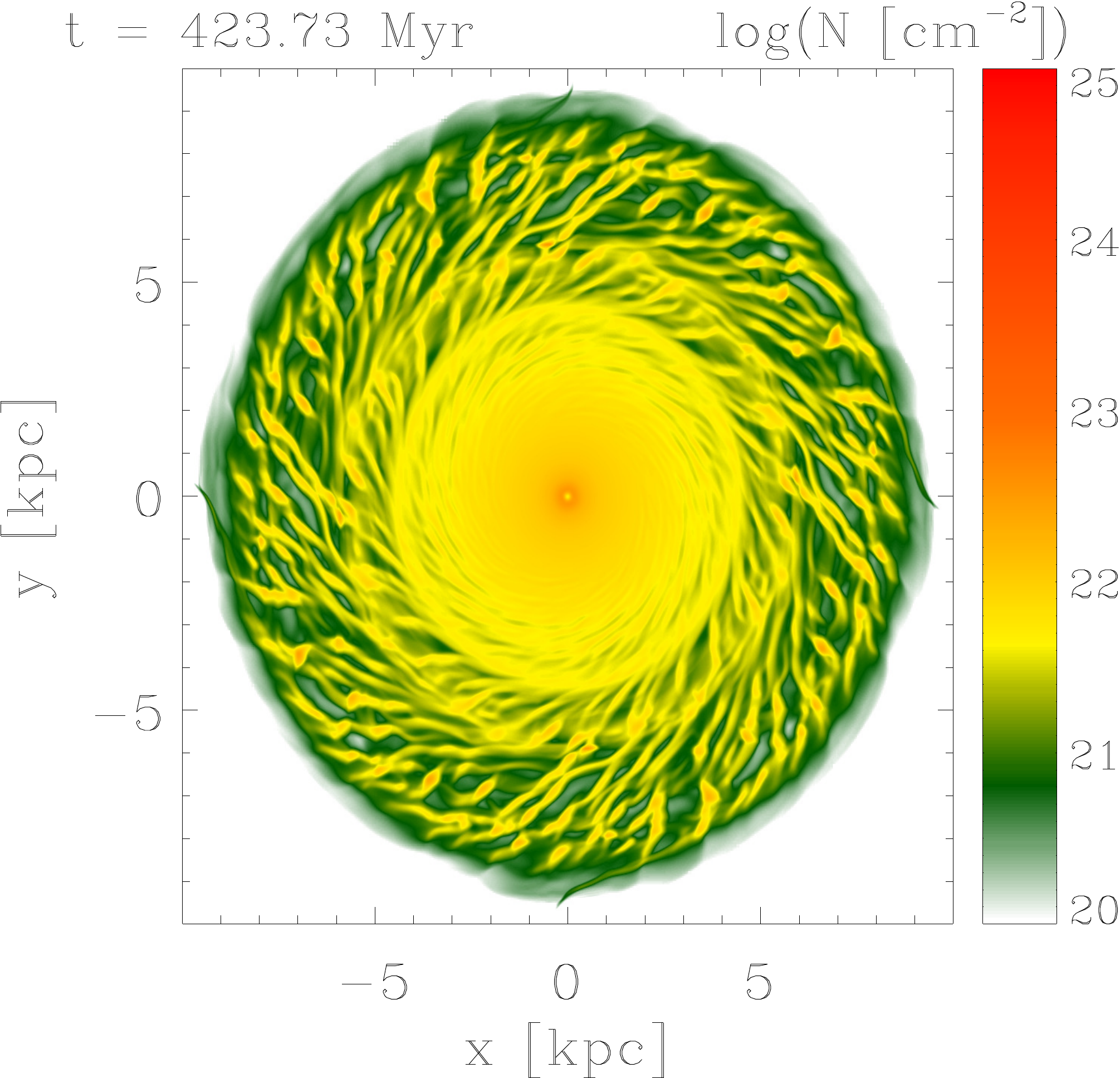}&\includegraphics[width=0.33\textwidth]{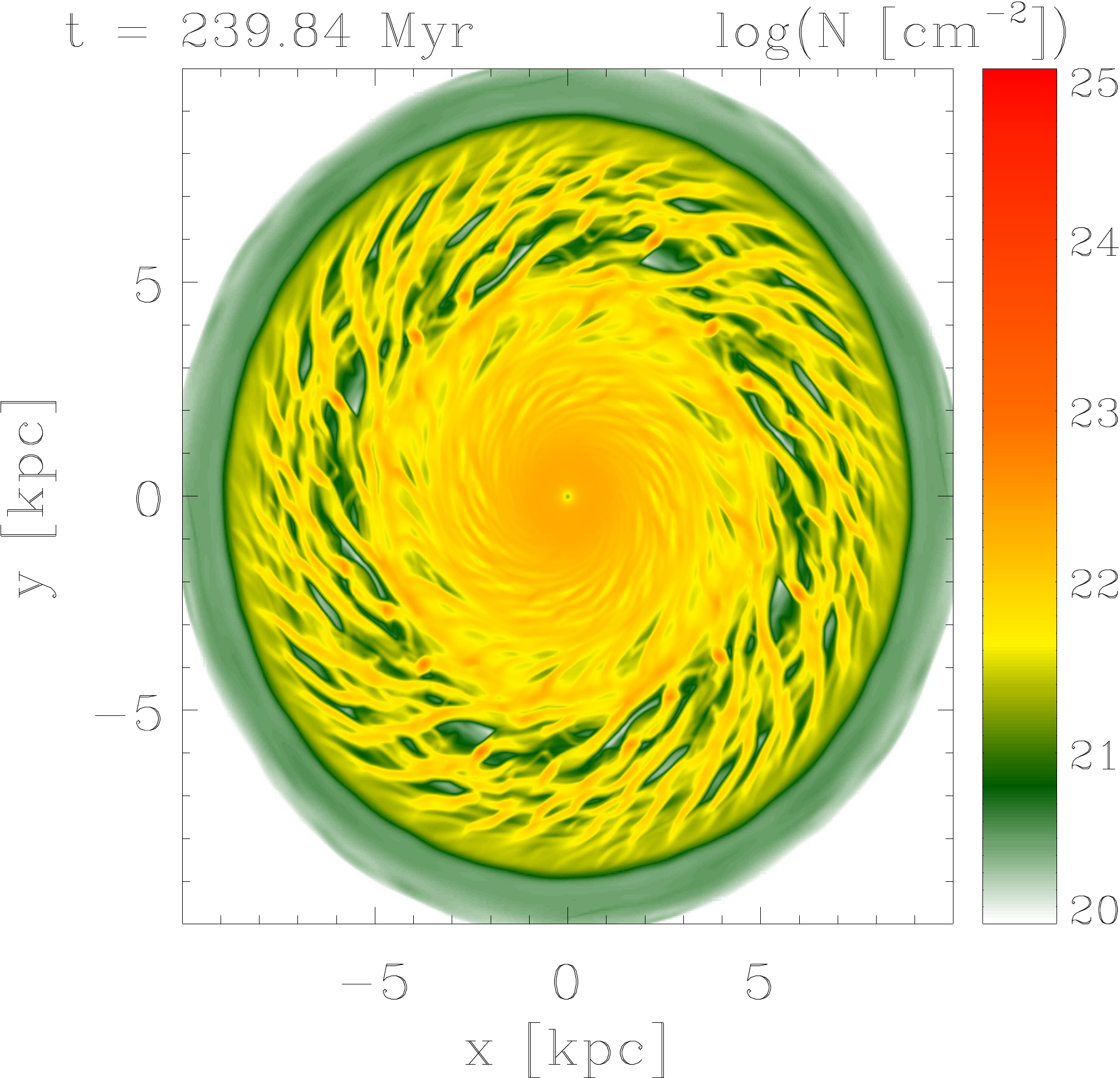}\\
   \includegraphics[width=0.33\textwidth]{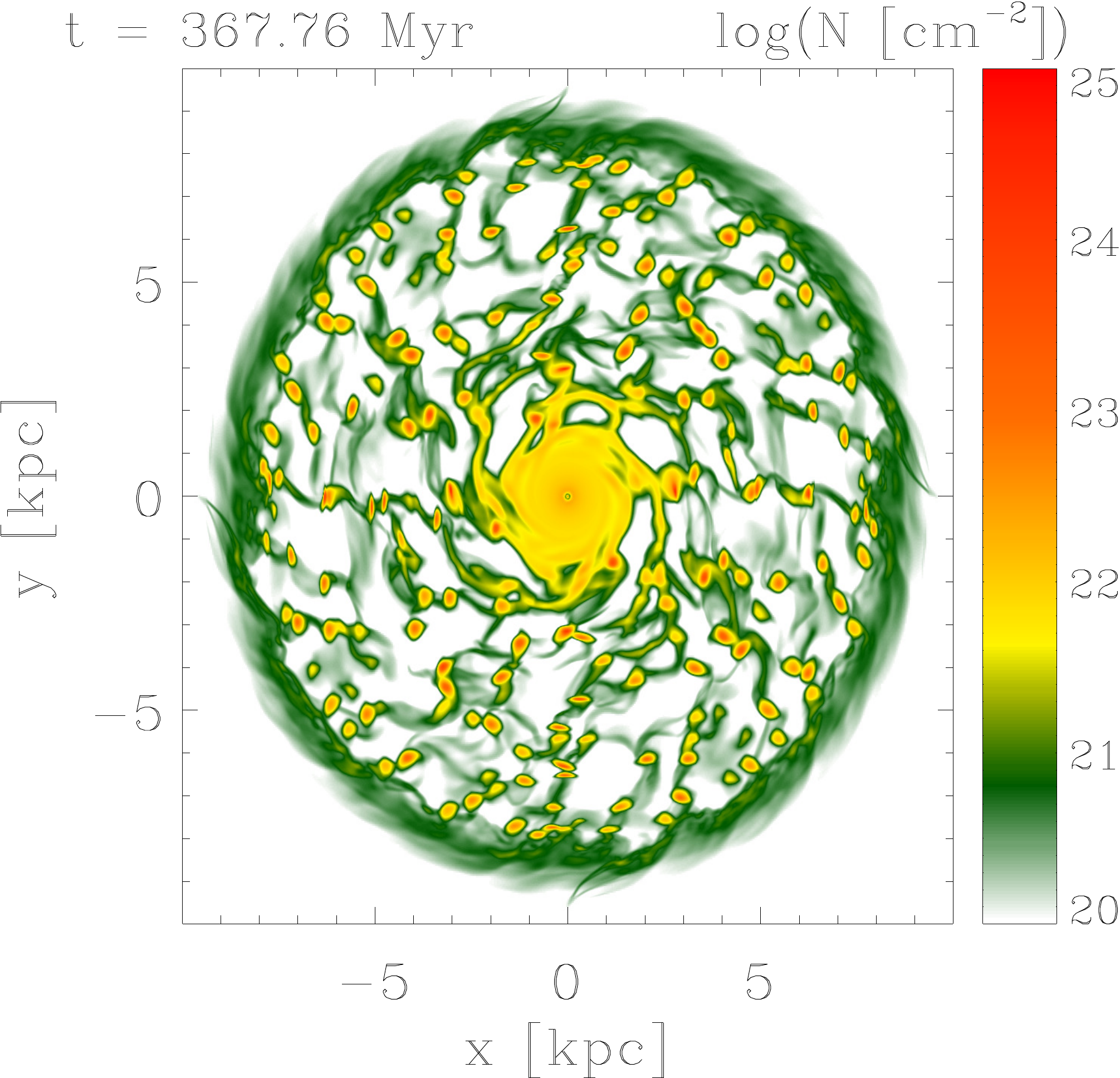}&\includegraphics[width=0.33\textwidth]{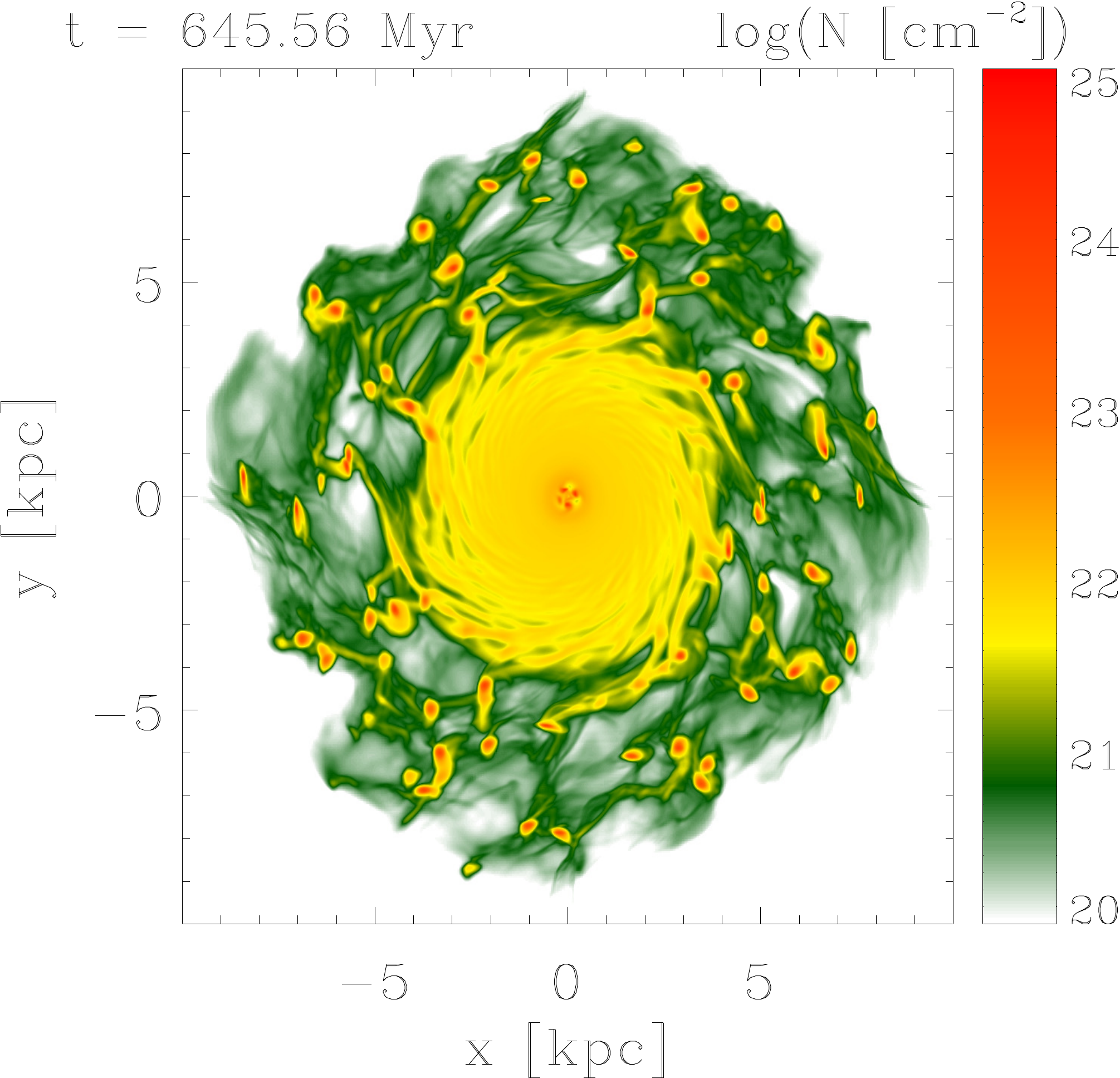}&\includegraphics[width=0.33\textwidth]{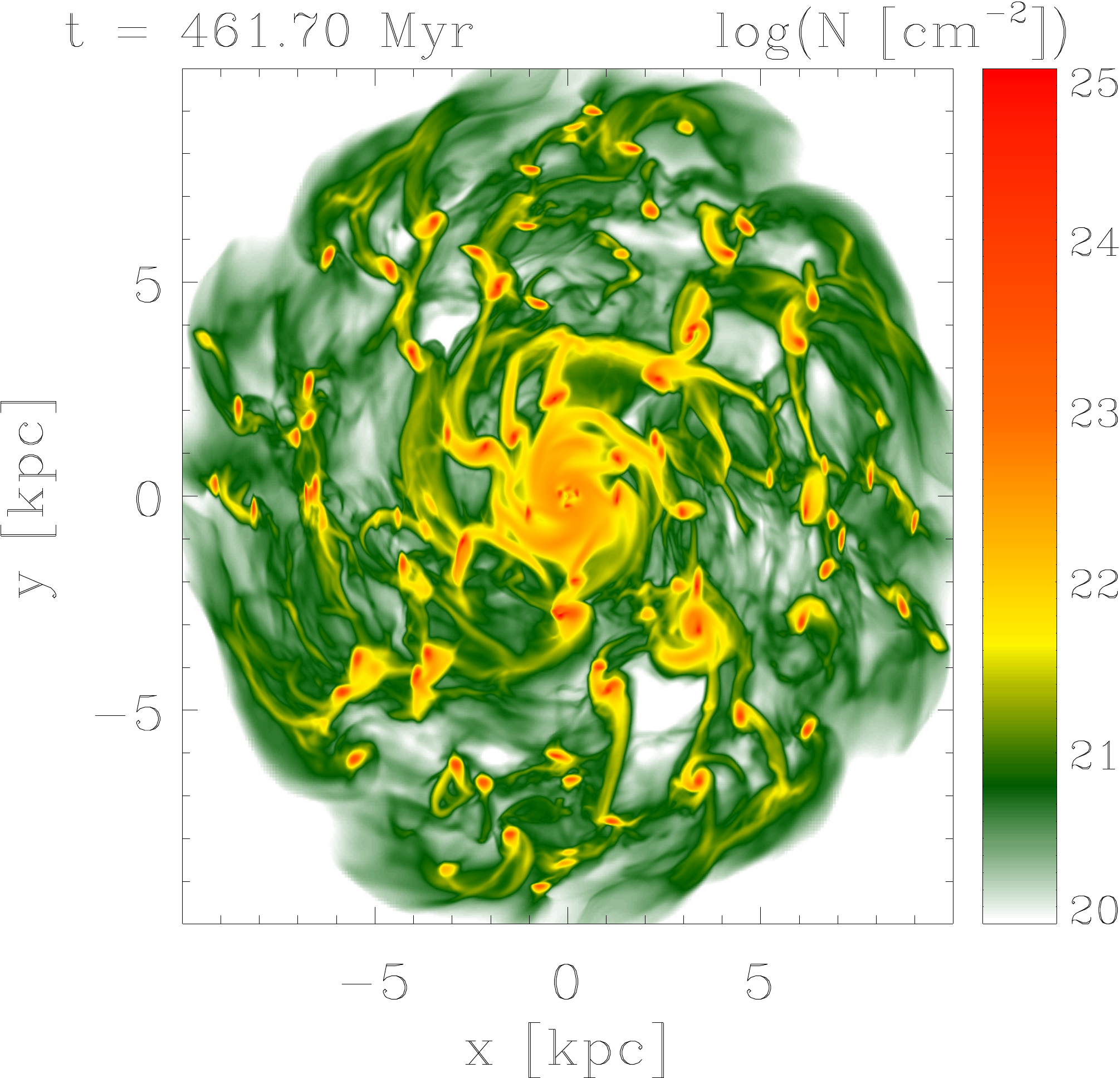}
  \end{tabular}
  \caption{Face-on column density maps of three galaxies at times $t_\mathrm{frag}$ (top) and $t_\mathrm{onerot}$ (bottom). It is evident that the magnetic field induces a different mode of fragmentation. 
  The long term evolution (bottom) shows that the diffuse gas appears to be supported by the field against becoming accreted onto the formed clouds. For more details, please see text.}
  \label{figColDensFO}
\end{figure*}

In Fig.~\ref{figColDensFO} we show face-on column density maps of galaxies with an initial plasma-$\beta=\left\{\infty,1,0.25\right\}$ at times $t_\mathrm{frag}$ and $t_\mathrm{onerot}$. Close inspection 
of this series of images reveals two crucial points. The first is the (above mentioned) mode of fragmentation. The hydrodynamic galaxy fragments into ring-like structures due to the radial Toomre 
instability. In contrast, the magnetised galaxies reveal a fragmentation pattern, which extends both in the azimuthal and radial direction \citep[see also][]{Koertgen18L}. This mode is a consequence of the 
Parker instability and its shorter dynamical timescale compared to the classical Toomre instability. It is further observed that the magnetised discs fragment at different galactocentric distances. The 
inner part out to $R\sim 5\,\mathrm{kpc}$ still reveals a smooth density distribution for the disc with $\beta_\mathrm{init}=1$. The stronger magnetised disc is in the process of fragmentation in this 
region. The spur-like features are clearly identified as overdense regions, where material accumulates in the magnetic valleys formed by the PI. At distances $R\sim6-7\,\mathrm{kpc}$, this disc has 
fragmented the most, while the weaker magnetised disc fragments more strongly towards $R\sim8-9\,\mathrm{kpc}$. This is due to the varying initial magnetisation of the disc, which implies that 
the conditions for an efficient PI are met at different galactocentric distances.\\
The second point is the \ita{time of fragmentation} (see also Table~\ref{tabFrag}). The hydrodynamic disc fragments first. Among the magnetised discs it is the one with highest magnetisation, which fragments the earliest. From 
the times denoted in Fig.~\ref{figColDensFO} and Table~\ref{tabFrag} it is evident that the disc with $\beta=0.25$ fragments almost 200\,Myr earlier than the disc with an equilibrium magnetic field, $\beta=1$. We will see 
later on that this trend extends towards higher $\beta$-values. At first sight, this behaviour appears counter-intuitive. Magnetic pressure acts to stabilise the gas against fragmentation \citep{Mestel56}. 
However, the magnetic field is Parker unstable with an associated timescale 
\beq
t_\mathrm{dyn,p}\sim H/v_\mathrm{a}\,\propto \sqrt{\beta}.
\eeq
 Following \citet{Kim02}, the typical timescale for gravitational contraction is 
 \beq
 t_\mathrm{grav}\sim \frac{\sqrt{c_\mathrm{s}^2+\frac{v_\mathrm{a}^2}{2}}}{G\Sigma_0}= \frac{c_\mathrm{s}\sqrt{1+\frac{1}{\beta}}}{G\Sigma_0}\propto\sqrt{1+\beta^{-1}}.
 \eeq
 Hence, the timescale for the PI \ita{decreases} with decreasing $\beta$, while the corresponding gravitational timescale 
 increases, and their ratio scales with $\sqrt{\beta}$.
 
\begin{figure*}
  \begin{tabular}{cc}
    \rotatebox{90}{\qquad\qquad \Large{\fat{Hydro}}}&\includegraphics[width=0.95\textwidth]{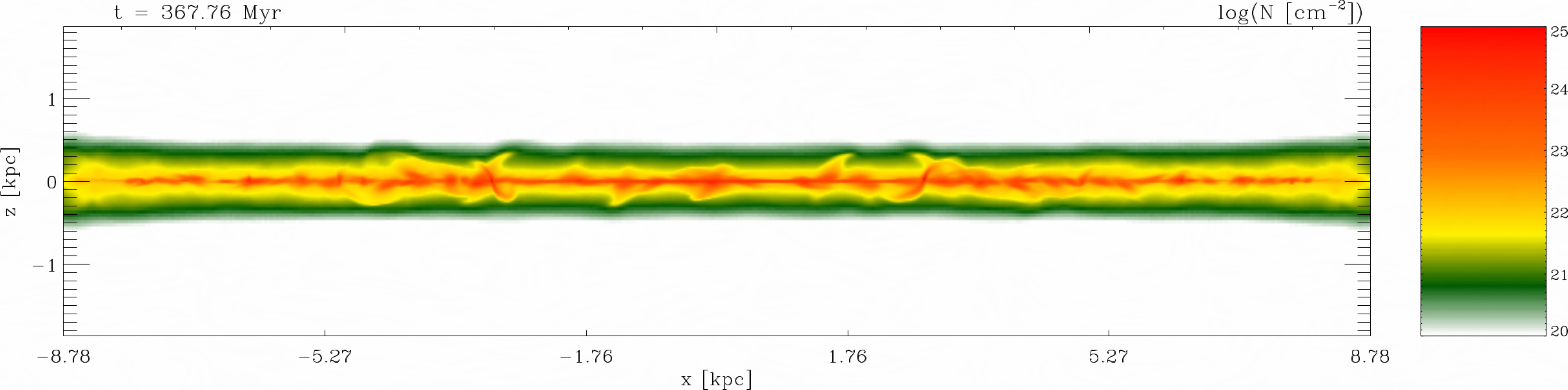}\\
   \rotatebox{90}{\qquad\qquad \Large{\fat{Beta1}}} &\includegraphics[width=0.95\textwidth]{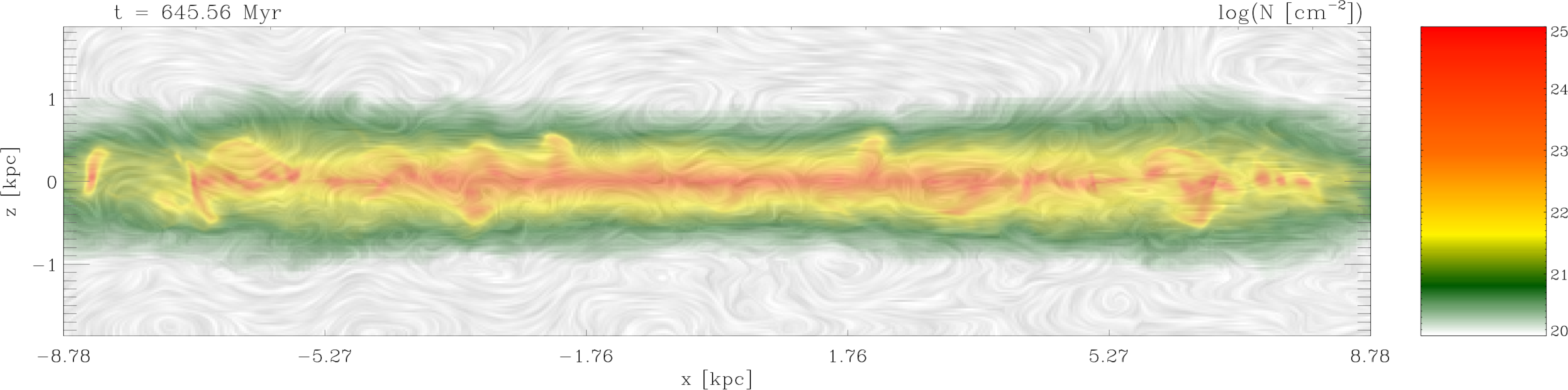}\\
   \rotatebox{90}{\qquad\qquad \Large{\fat{Beta0.25}}}&\includegraphics[width=0.95\textwidth]{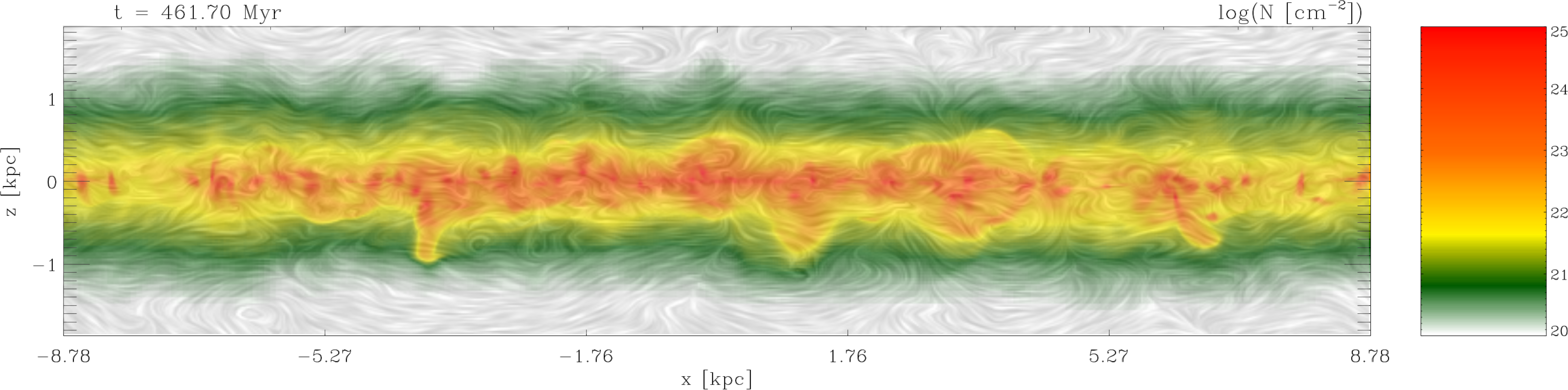}
  \end{tabular}
  \caption{Edge-on column density and magnetic field line maps of three fiducial galaxies at time $t_\mathrm{onerot}$. The drapery pattern in the middle and bottom panels depicts the magnetic field morphology as retrieved from a 
  line integral convolution (based on the IDL routine written by D.~Falceta Gon\c{c}alves). Note the increased disc height with increasing magnetisation. Note the difference in the drapery patterns. 
  While the pattern appears smooth for the disc with $\beta=1$, it reveals a generally more tangled/perturbed field morphology for the much more fragmented disc with $\beta=0.25$. This is especially clear for regions with an  
  enhanced number of clouds. Here, the field changes its orientation as a consequence of the varying molecular cloud densities \citep[see also][]{Planck16c}.}
  \label{figColDensEO}
\end{figure*}

\begin{figure*}
	\begin{tabular}{cc}
		\fat{\Large{Beta1}}		&\fat{\Large{Beta0.25}}\\
		\includegraphics[width=0.4\textwidth]{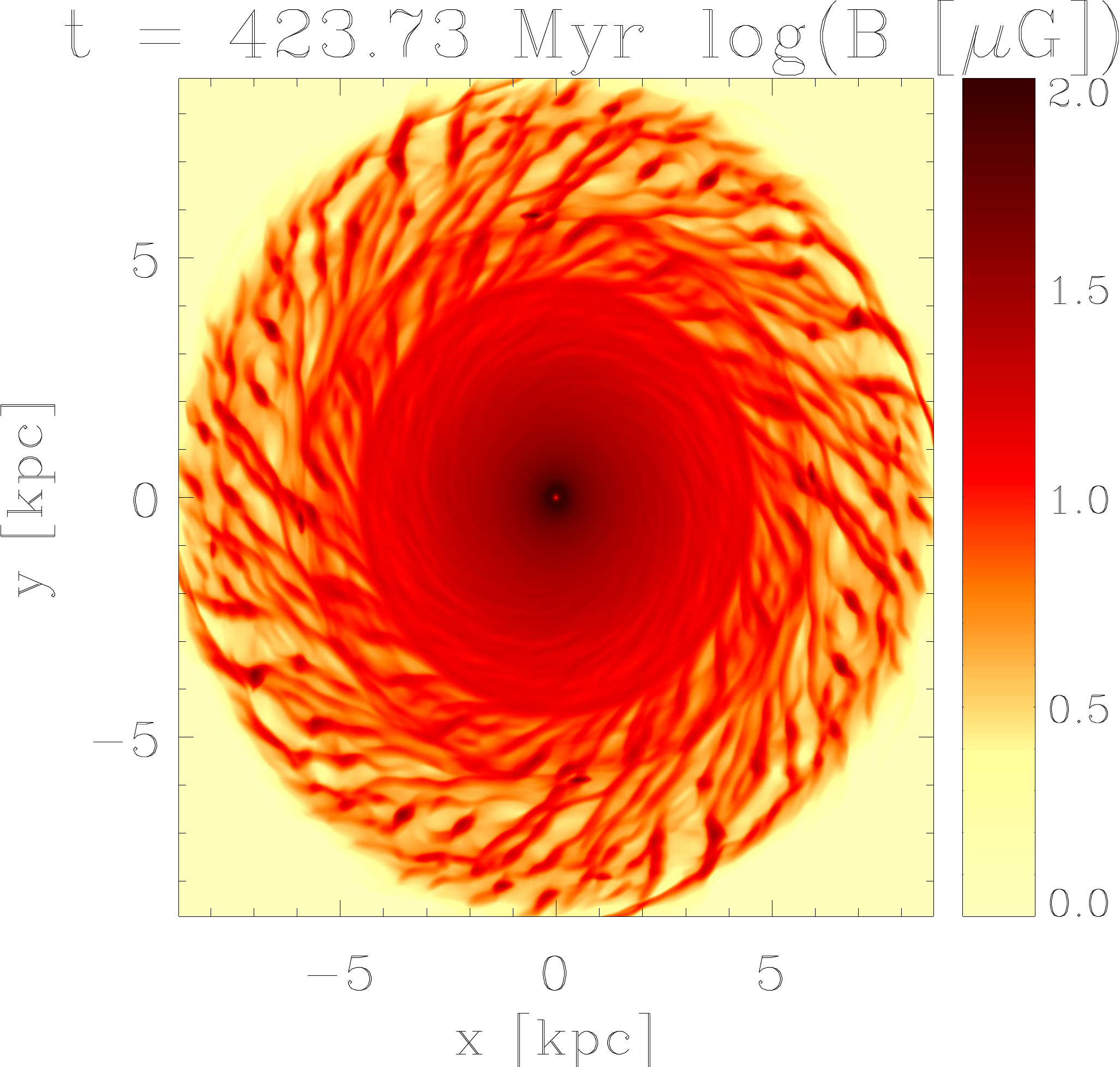}	&\includegraphics[width=0.4\textwidth]{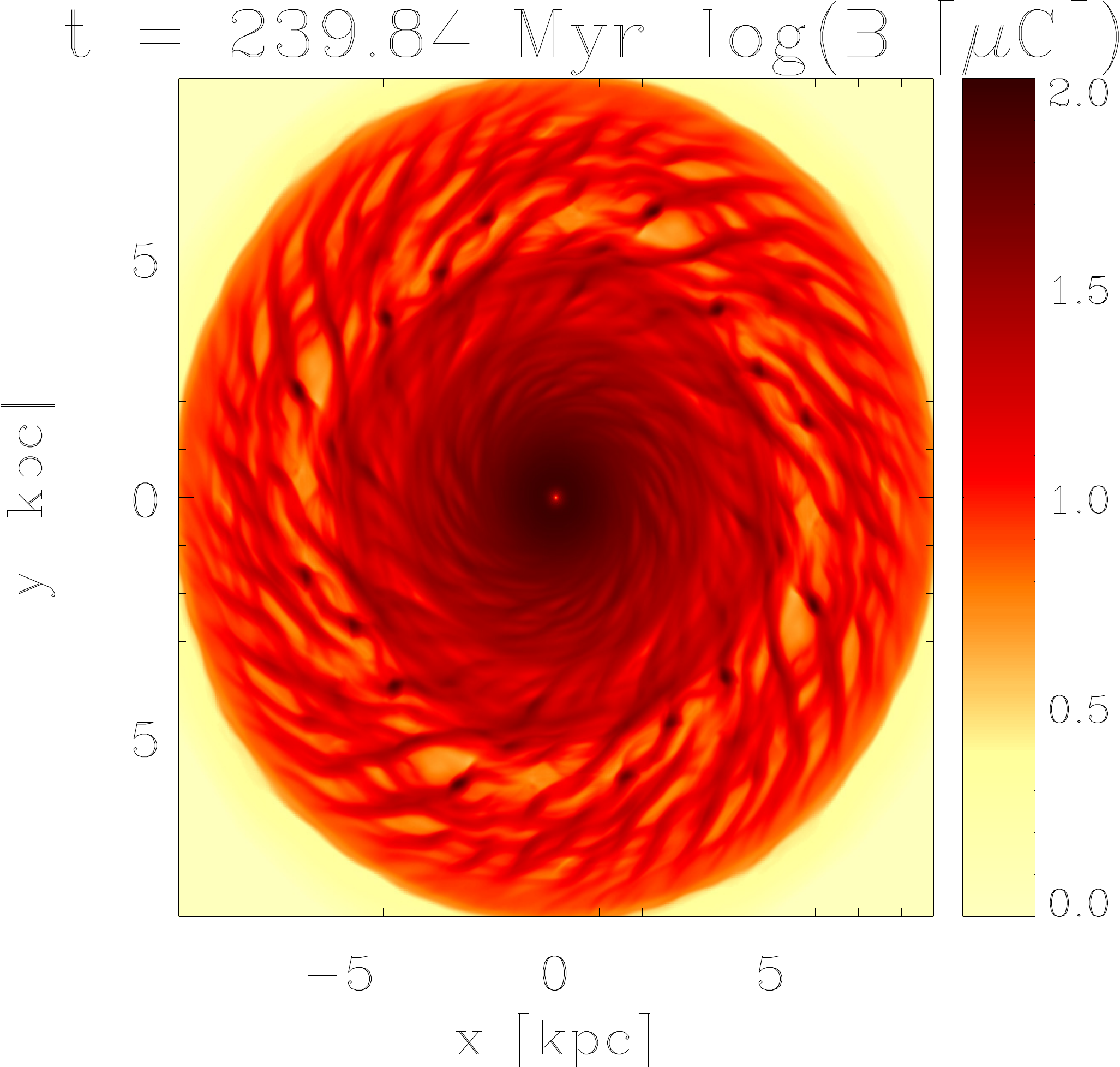}\\
		\includegraphics[width=0.4\textwidth]{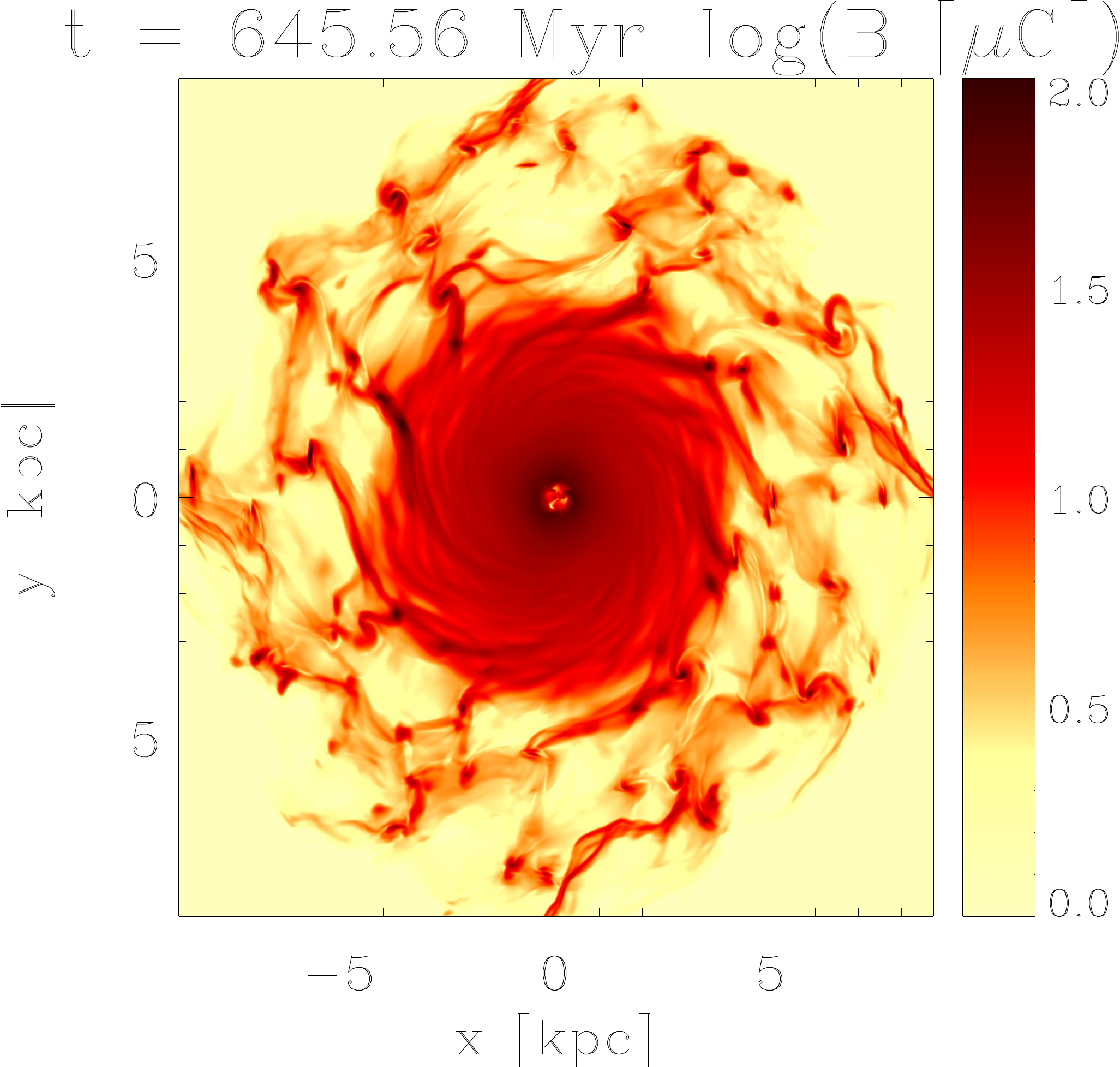}	&\includegraphics[width=0.4\textwidth]{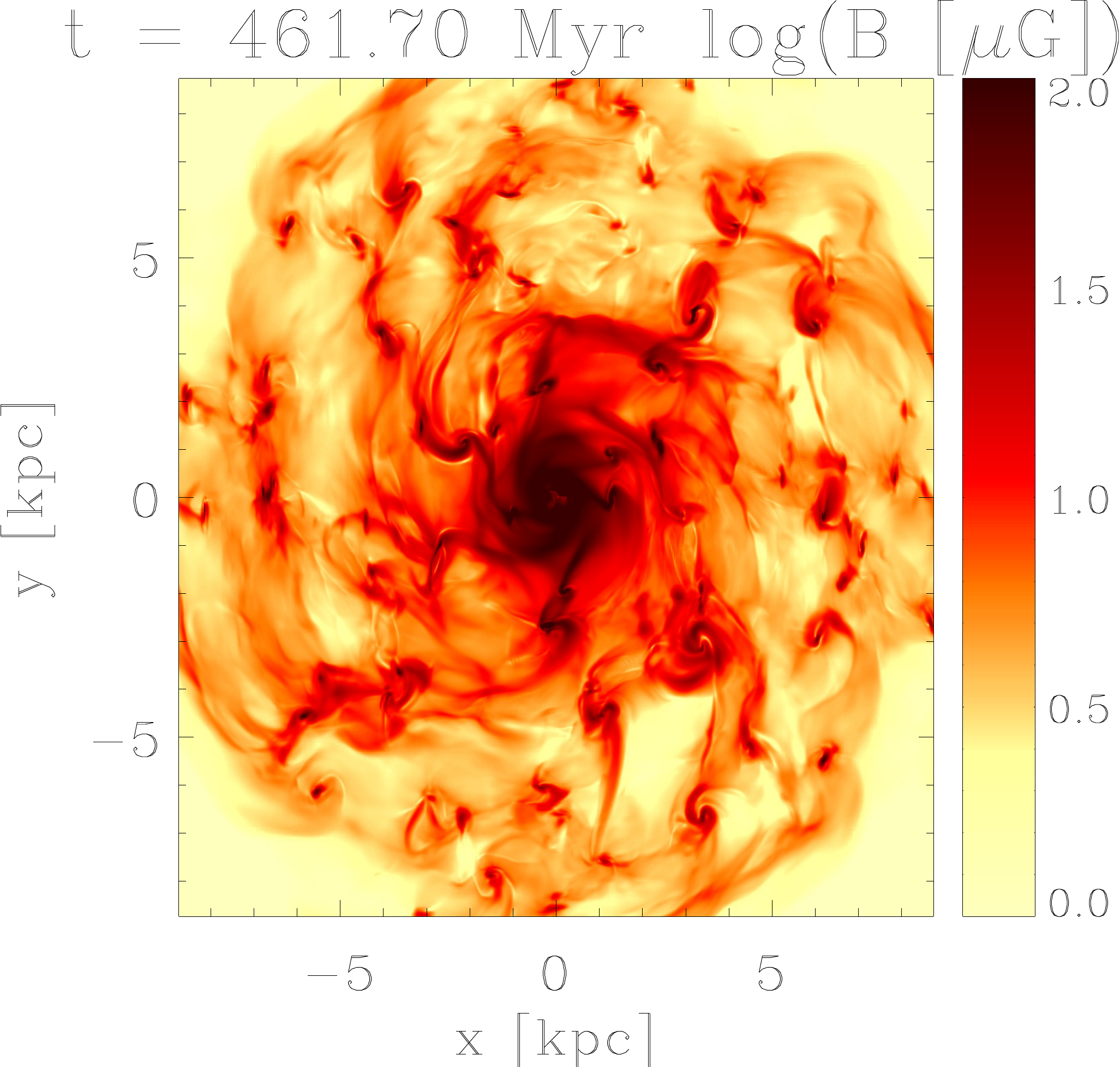}\\
	\end{tabular}
	\caption{Maps of the density weighted magnetic field strength of the discs with $\beta=1$ (left) and $\beta=0.25$ (right) 
	at $t=t_\mathrm{frag}$ (top) and $t=t_\mathrm{onerot}$ (bottom). At the time of fragmentation into individual clouds, 
	the filamentary structures due to the Parker instability can still be recognised. After one additional rotation, large 
	magnetic structures are identified. The magnetisation of the formed clouds appears similar in both discs.}
	\label{figTotMag}
\end{figure*}

With increasing $\beta$, the Parker timescale increases, thereby delaying the onset of the instability. However, when the ratio 
of thermal to magnetic energy is significantly larger than unity, magnetic support is weak and the disc fragments earlier (due to gravitational instability) compared to discs in equilibrium (see e.g. CMM phase in Fig.~\ref{figTimeMass}), 
preferably due to a combination of variations in the gas surface density and the plasma-$\beta$.\\
We further show in Fig.~\ref{figColDensFO} the discs at time $t_\mathrm{onerot}$. At this time all large scale structures have broken up into individual clouds. All galaxies show a rather smooth density 
pattern in the center, which has not fragmented, yet. The extent of this region is larger for the disc with $\beta=1$, indicating that fragmentation proceeds slower here.\\
The fragmented parts in the discs reveal marked differences. At first, the number of clouds in the hydrodynamic disc is much larger compared with the MHD discs. Secondly, clouds in the magnetised discs 
appear larger, since the magnetic field balances gravitational contraction. In addition, their inter-cloud medium is composed of filamentary structures. In the hydrodynamic disc, such features are only 
observed in regions, where two or more clouds closely encounter each other. This is not an indication for more frequent cloud-cloud interactions in the magnetised discs. It rather shows that the diffuse 
gas is supported by the magnetic field against accretion onto the clouds. Please note further the increased disc size of the disc with $\beta=0.25$.\\
In Fig.~\ref{figColDensEO}, we show edge-on column density maps as well as the magnetic field structure of the three discussed discs. As expected, the higher the magnetic field strength, the larger the disc height due to the general 
magnetic buoyancy. For the 
selected discs, the differences in the vertical extent of the disc can be as large as a factor of $\sim5$ close to the center. Near the disc outskirts, the difference in disc height become smaller due to the 
vanishing influence of the magnetic field. Interestingly, in the case of the disc with the strongest magnetic field, the vertical extent of the gas appears to be independent of galactocentric distance.\\
While the hydrodynamic disc appears smooth in an edge-on view, the magnetised discs reveal much more substructure. Individual objects can be clearly identified in the latter discs, while the hydrodynamic one 
only reveals some localised features above/below the midplane. Similar features are also observed in the magnetised discs, but their extent is much broader and higher. However, these features come primarily 
with the disc fragmentation, which explains the differences between the two magnetised cases. \\
We further show the magnetic field structure, obtained from a line integral convolution, in the middle and bottom panel of Fig.~\ref{figColDensEO}. The magnetic field morphology reveals 
some degree of structure in the gas surrounding the galaxies. These structures are mainly vortex- or wave-like, where the wave-like pattern is indicative of a still active Parker instability of the field at higher latitudes. Apart from these 
features, the magnetic field is oriented mostly parallel to the column density iso-contours (or perpendicular to the column density gradient), as is typical for the low column density gas \citep[e.g.][]{Soler13,Planck16c}. Near the 
disc midplane, the picture gets more complicated due to the highly dynamic environment. Although the initial toroidal field component can be inferred, the field is strongly perturbed in regions, where multiple clouds are 
observed. \\
For completeness, Fig.~\ref{figTotMag} shows the density weighted magnetic field strength at the above defined 
temporal stages. The general increase in field strength due to the variation of the initial $\beta$ is readily seen. However, 
the similar mode of fragmentation is also revealed in these maps, and already formed clouds are recognised as strongly 
magnetised, almost spherical regions. \\
At later times, i.e. $t=t_\mathrm{onerot}$, major differences in the disc magnetisation are only seen in the more diffuse 
regions. The densest parts reveal similar values of the field strength, as they seem to decouple from the diffuse 
environment. The filamentary structures in the magnetic field strength maps are remnants of the large scale filaments, 
which were formed by the PI.
A detailled analysis of the orientation between the field and the (column-) density structures is postponed to a future study.

\subsection{The multiphase view of the discs}
\begin{figure*}
 \begin{tabular}{cccc}
 &\fat{WNM}	&\fat{CNM}	&\fat{CMM}\\
  \rotatebox{90}{\qquad\qquad\qquad\qquad \fat{\Large{Hydro}}}&\includegraphics[width=0.32\textwidth]{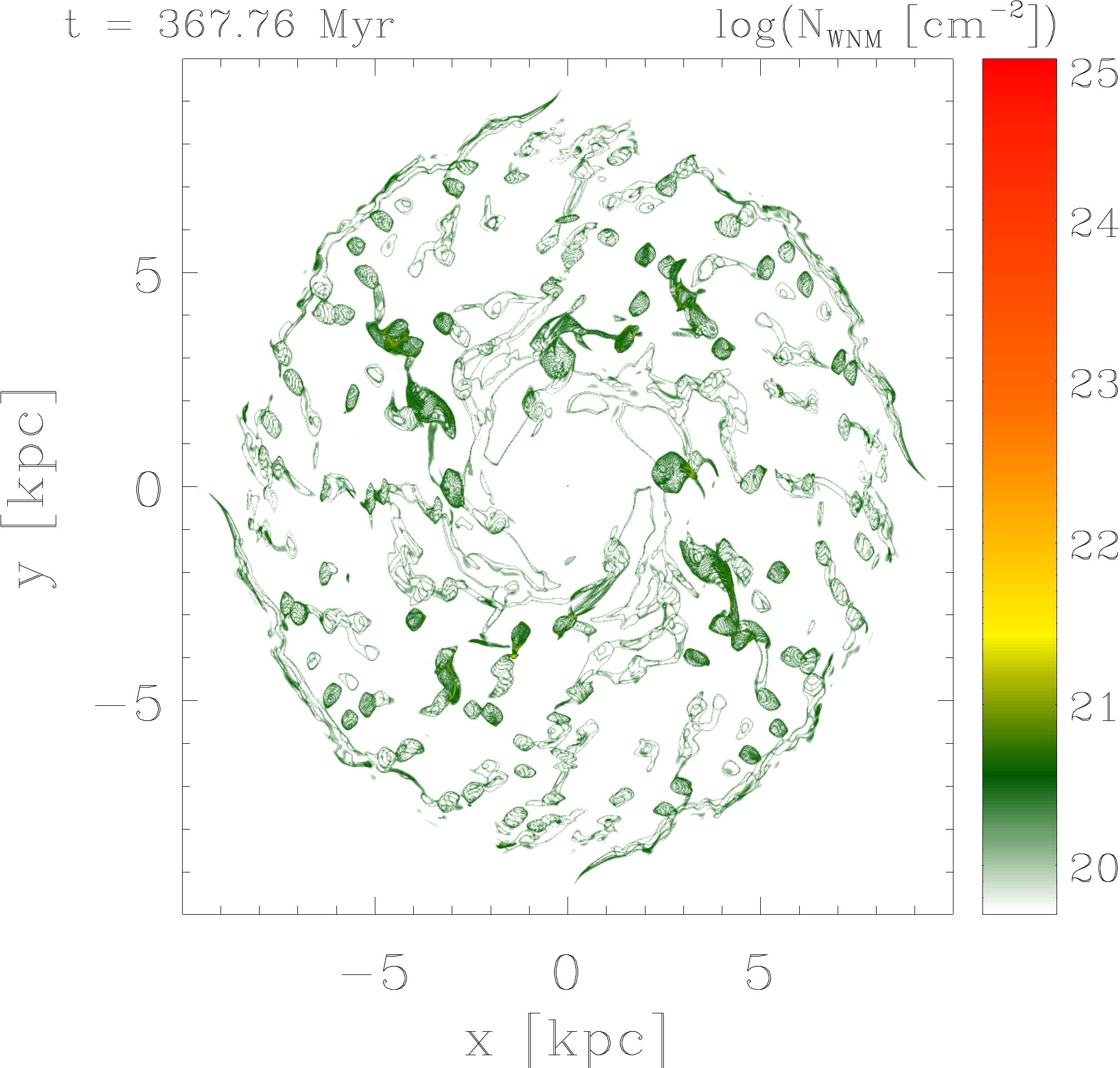}&\includegraphics[width=0.32\textwidth]{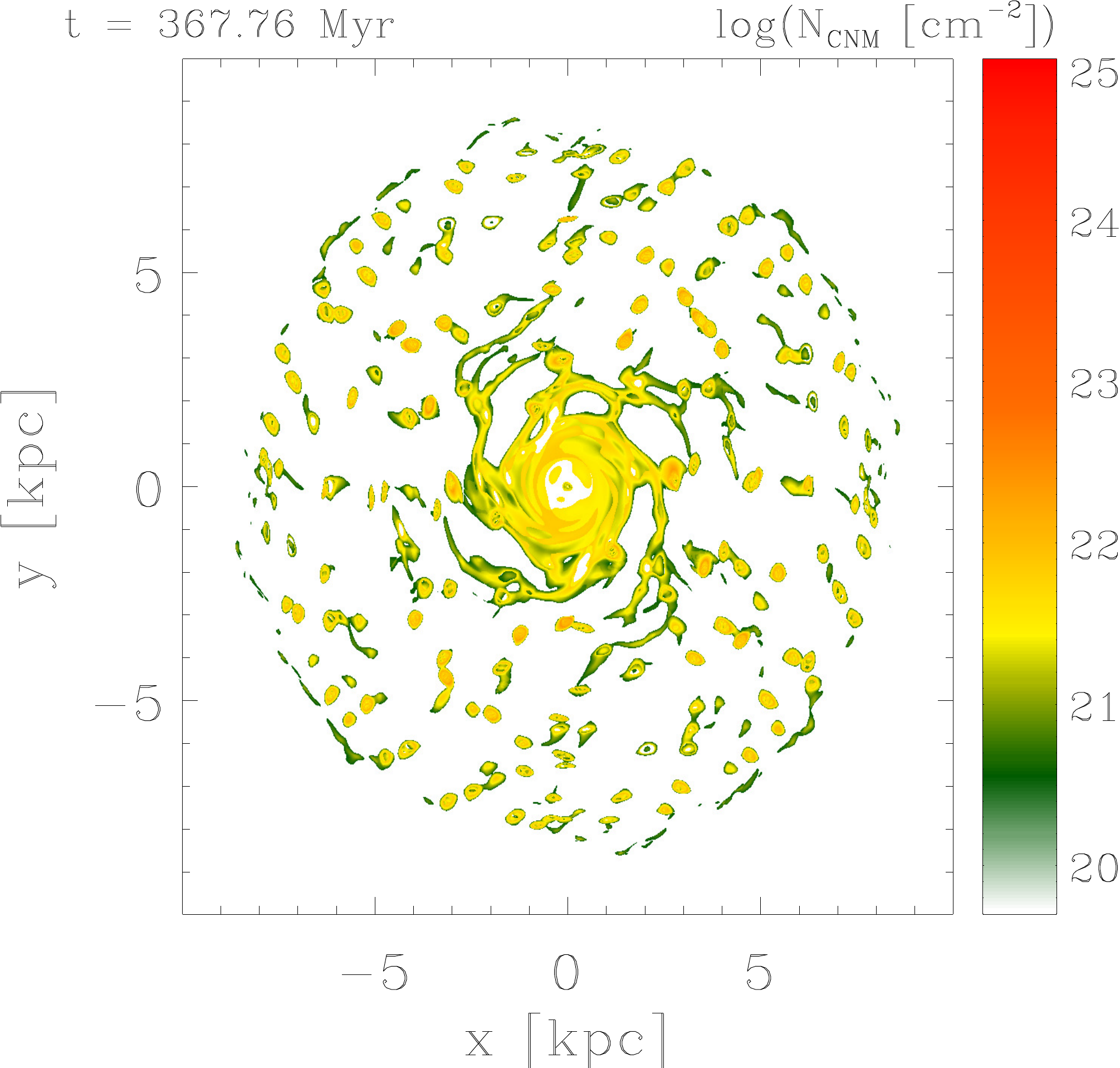}&\includegraphics[width=0.32\textwidth]{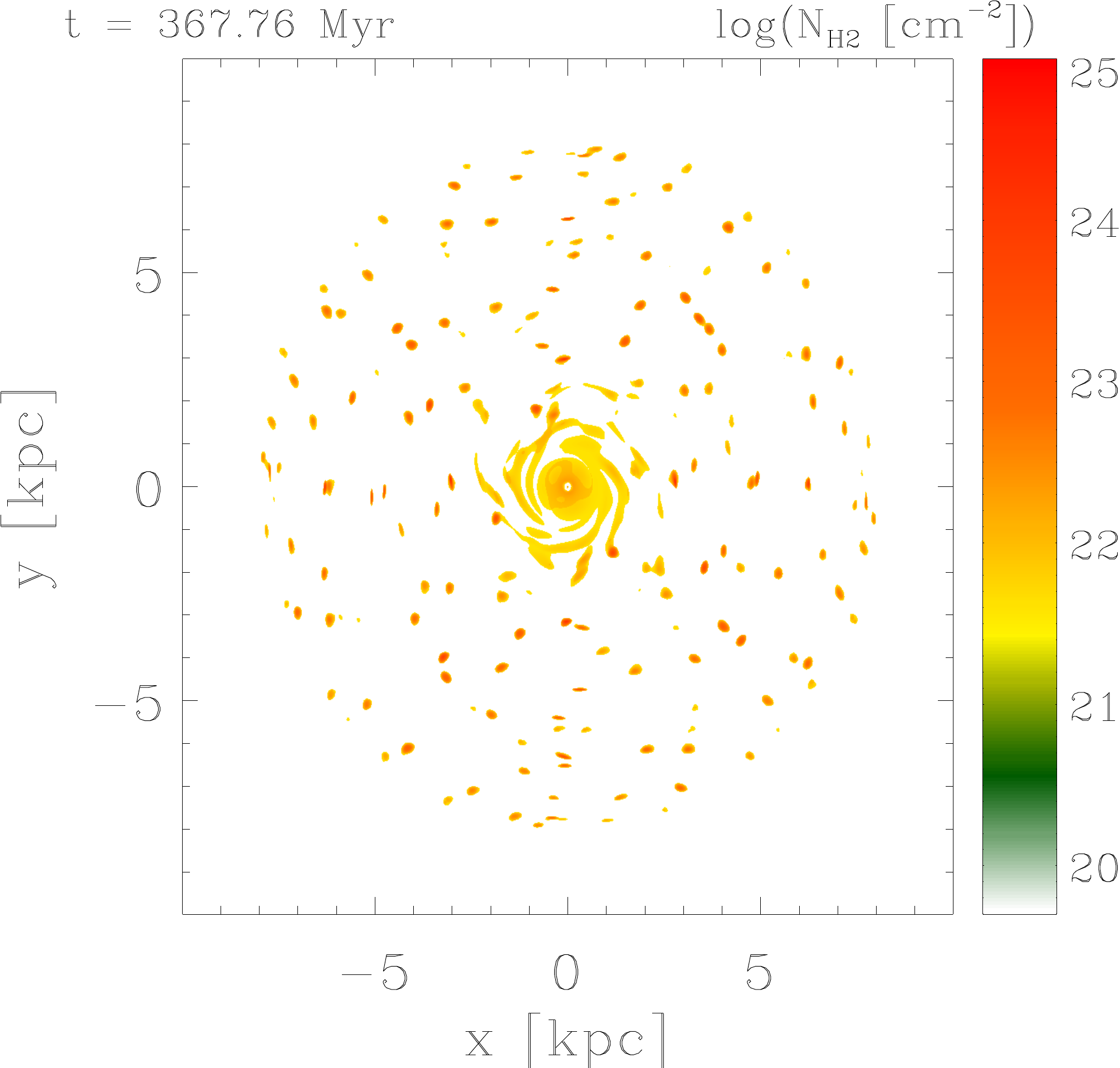}\\
  \rotatebox{90}{\qquad\qquad\qquad\qquad \fat{\Large{Beta1}}} &\includegraphics[width=0.32\textwidth]{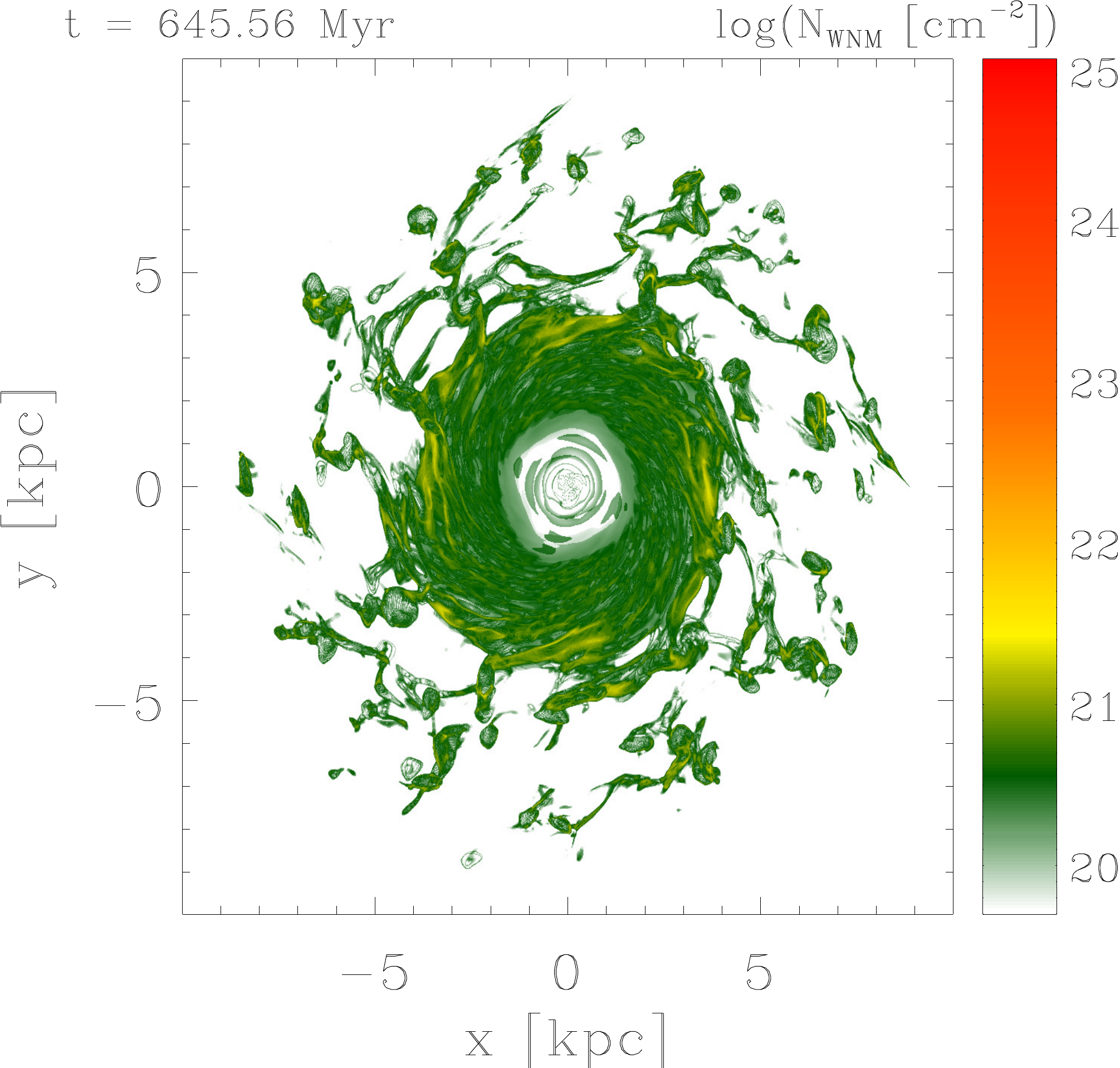}&\includegraphics[width=0.32\textwidth]{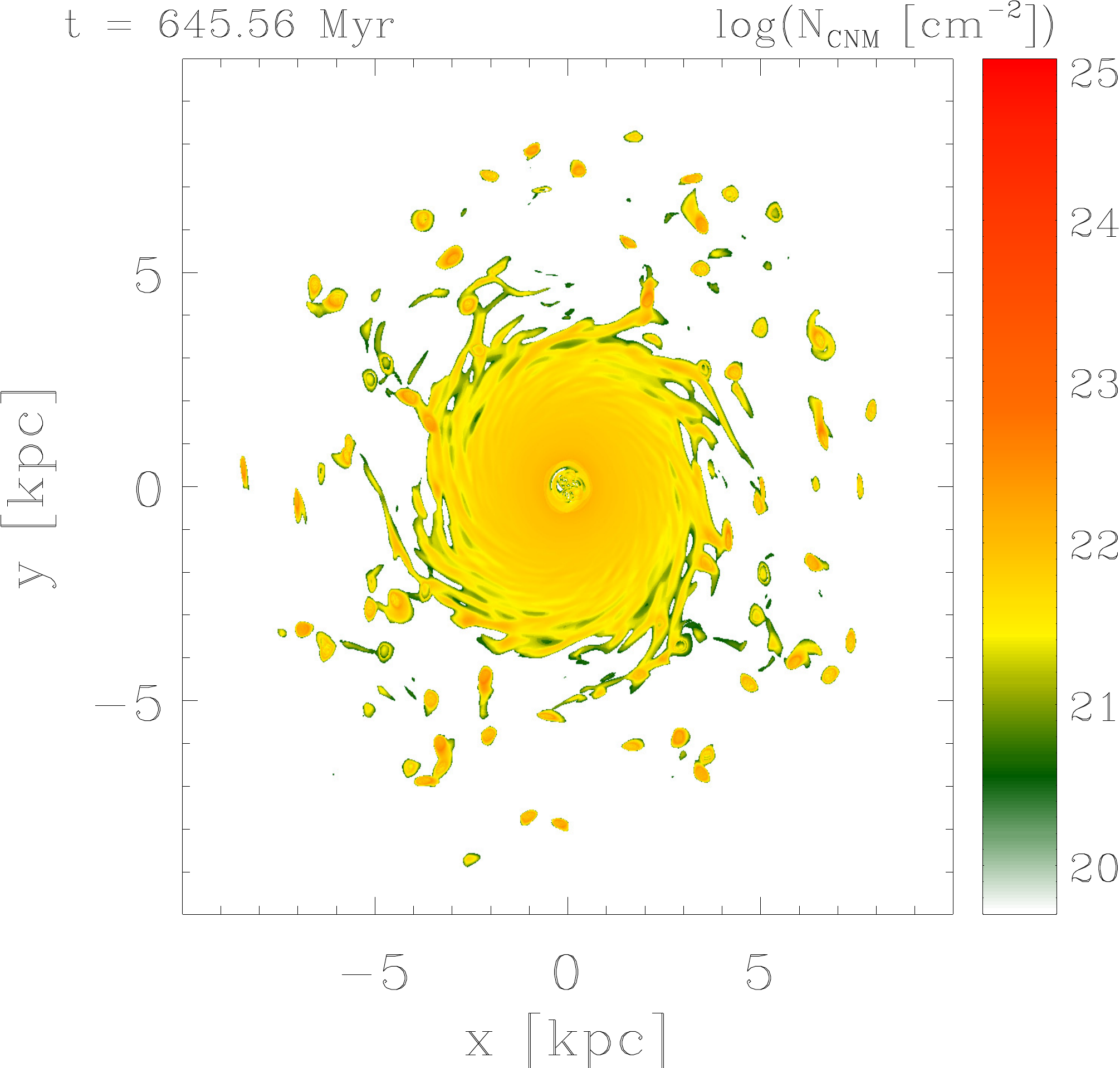}&\includegraphics[width=0.32\textwidth]{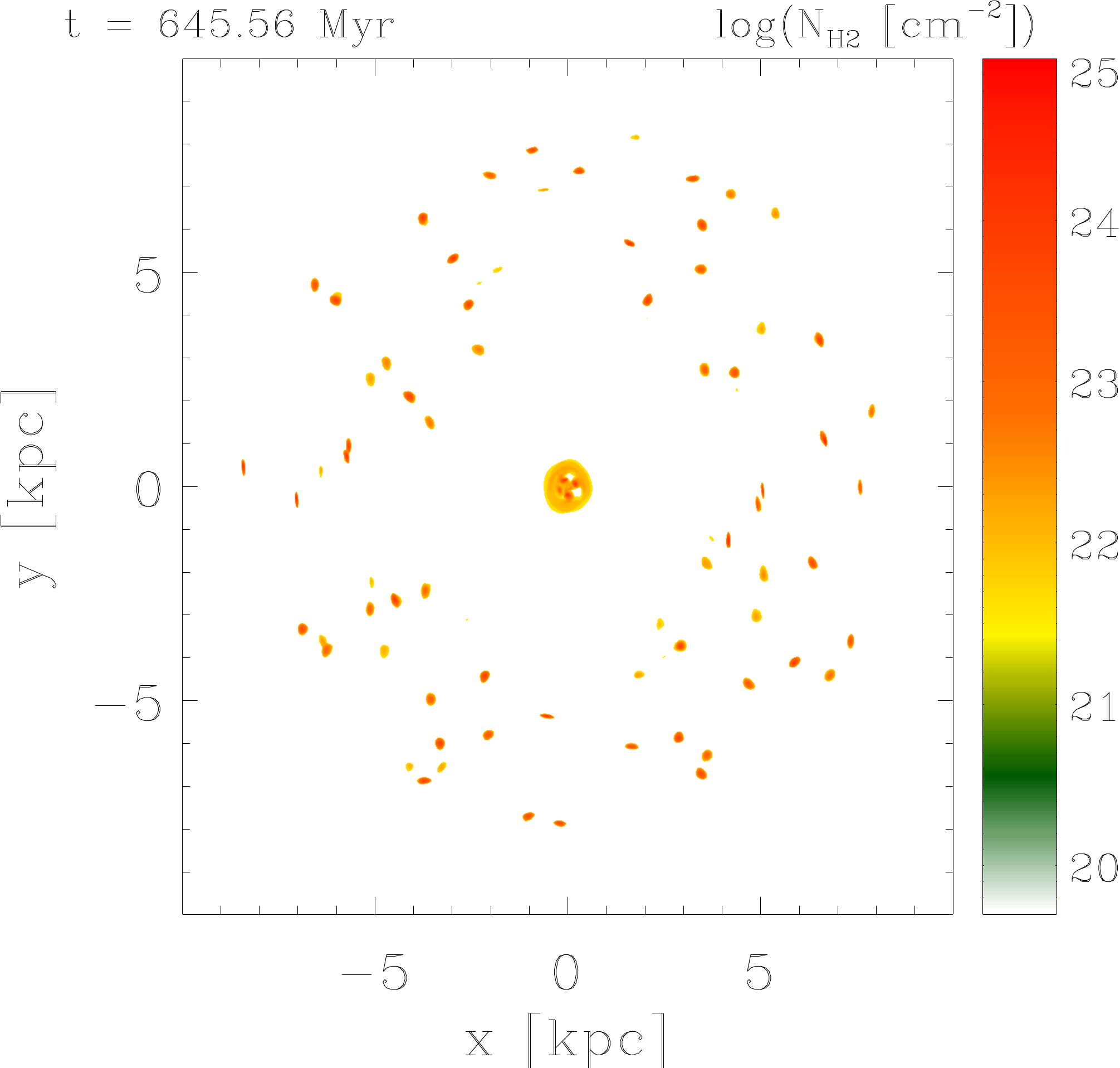}\\
    \rotatebox{90}{\qquad\qquad\qquad\qquad \fat{\Large{Beta0.25}}}&\includegraphics[width=0.32\textwidth]{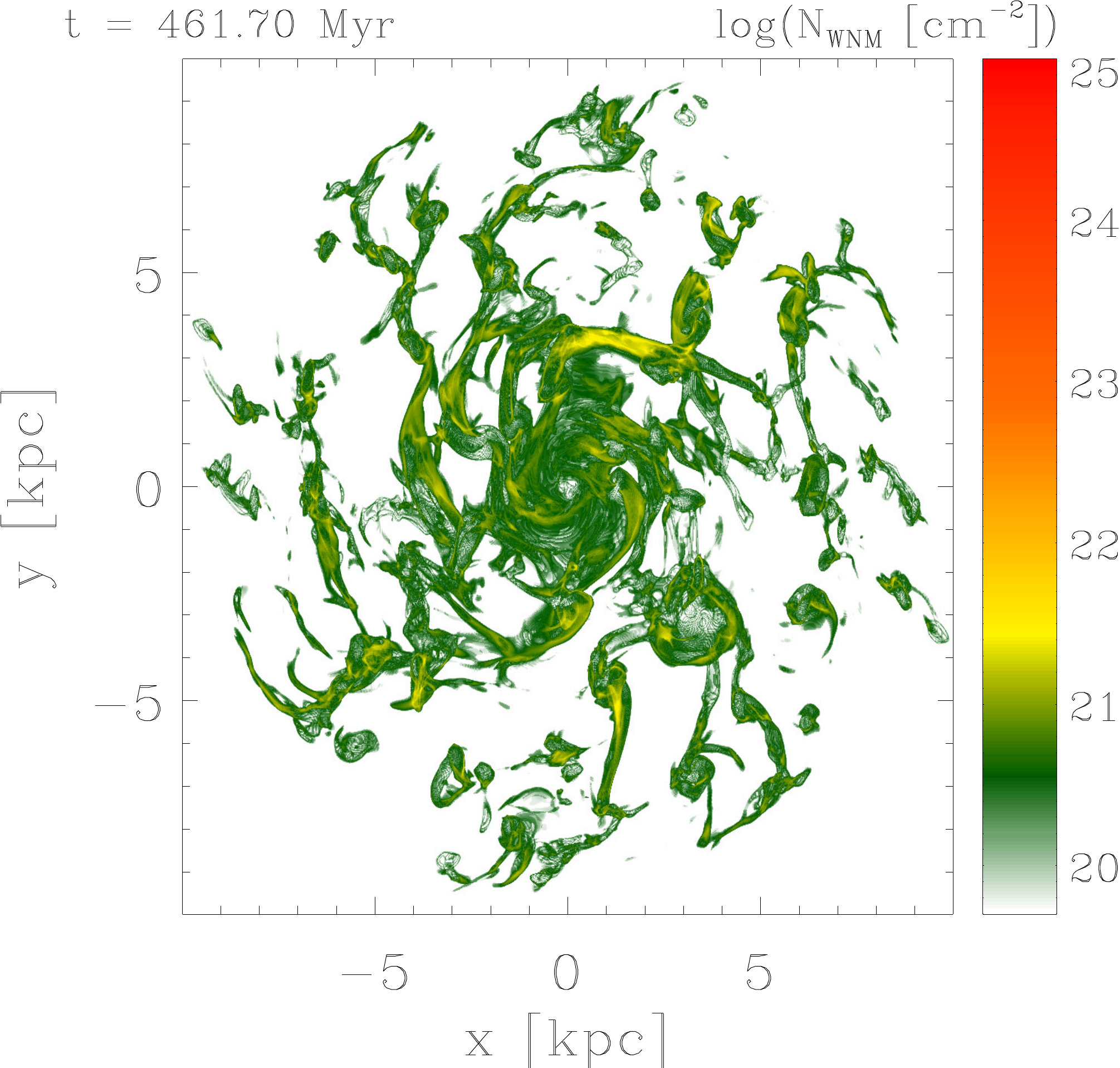}&\includegraphics[width=0.32\textwidth]{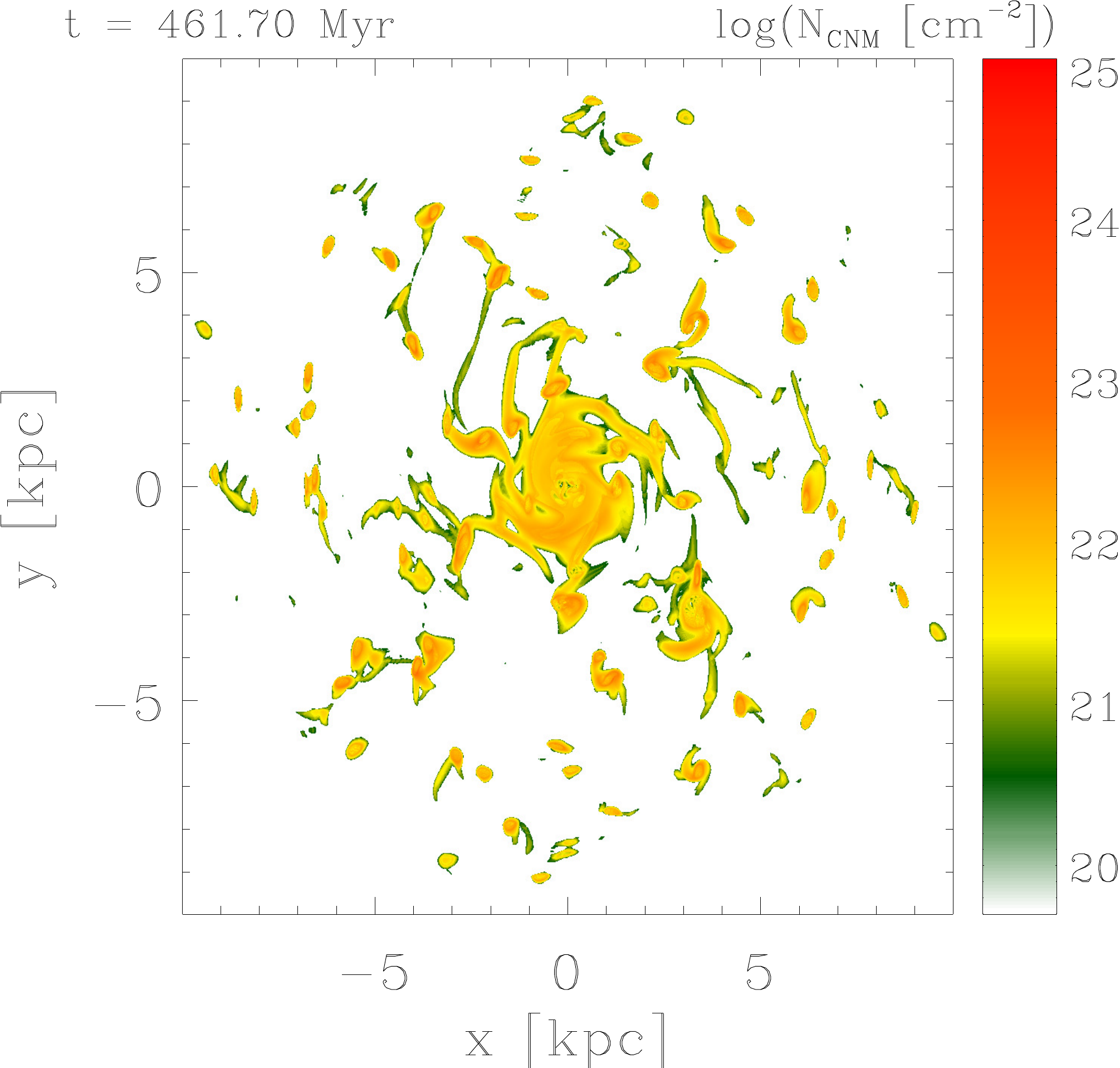}&\includegraphics[width=0.32\textwidth]{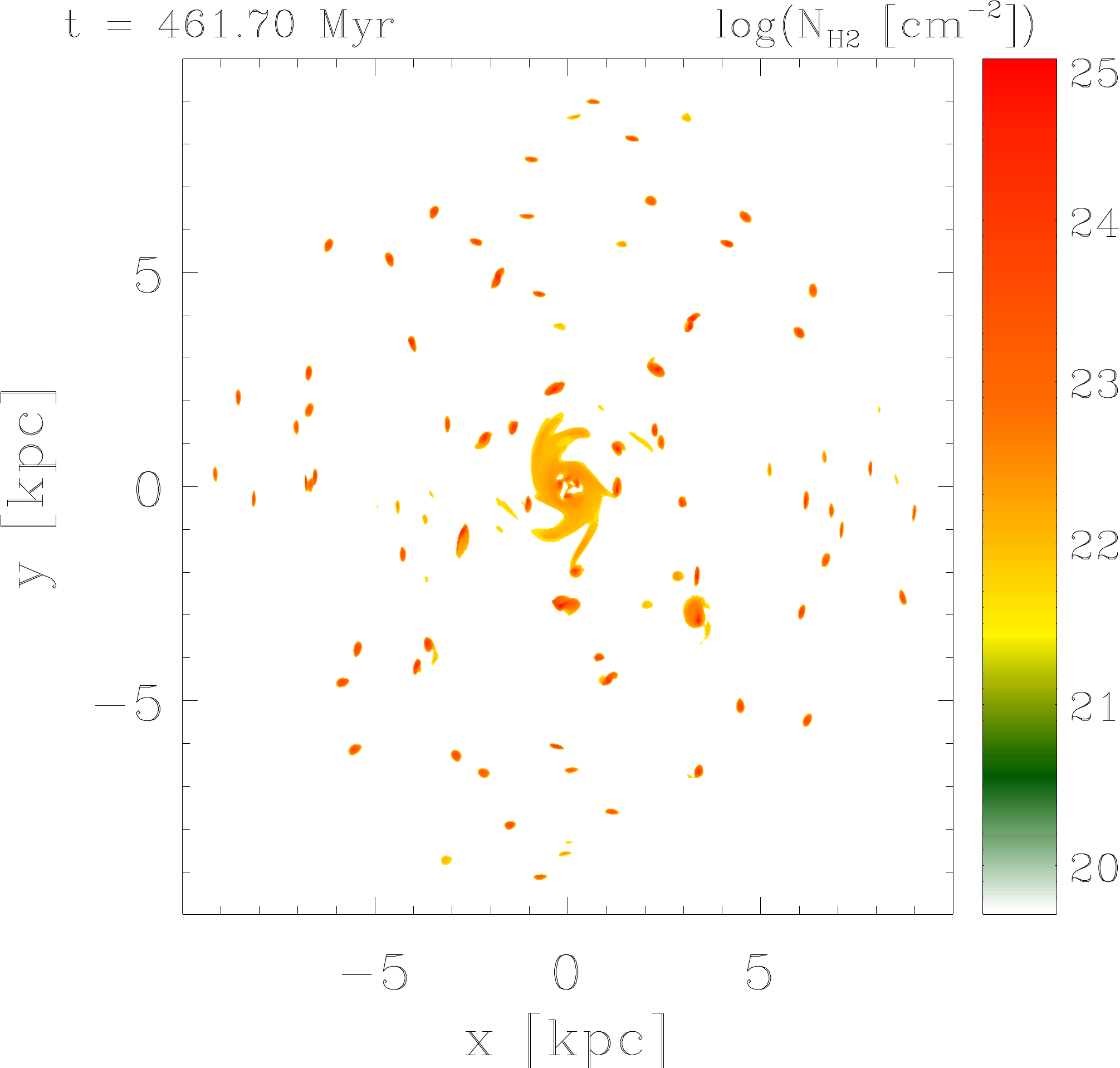}\\
 \end{tabular}
\caption{Column density maps of the different gas phases at time $t_\mathrm{onerot}$. Note the change in 
morphology of the concentrations of H\,I gas with increasing magnetic field strength, from being mostly spherical to almost filamentary.}
\label{figCdensFO}
\end{figure*}

\begin{figure*}
 \begin{tabular}{cc}
  \rotatebox{90}{\qquad\qquad \fat{Hydro}}&\includegraphics[width=0.95\textwidth]{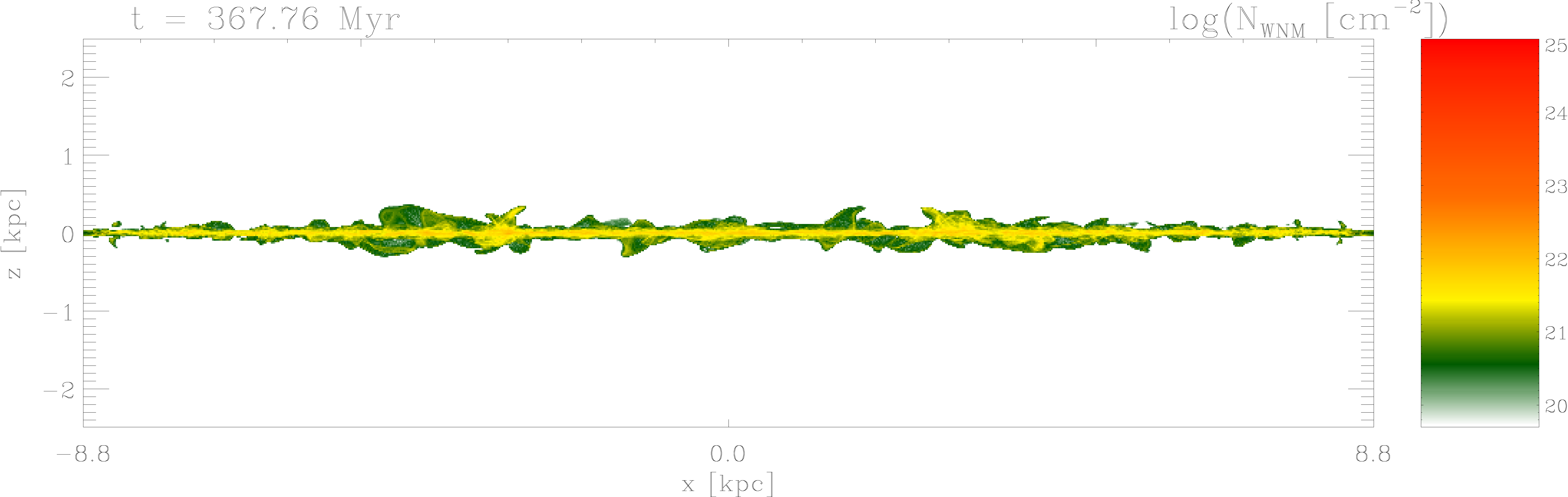}\\
  \rotatebox{90}{\qquad\qquad \fat{Beta1}} &\includegraphics[width=0.95\textwidth]{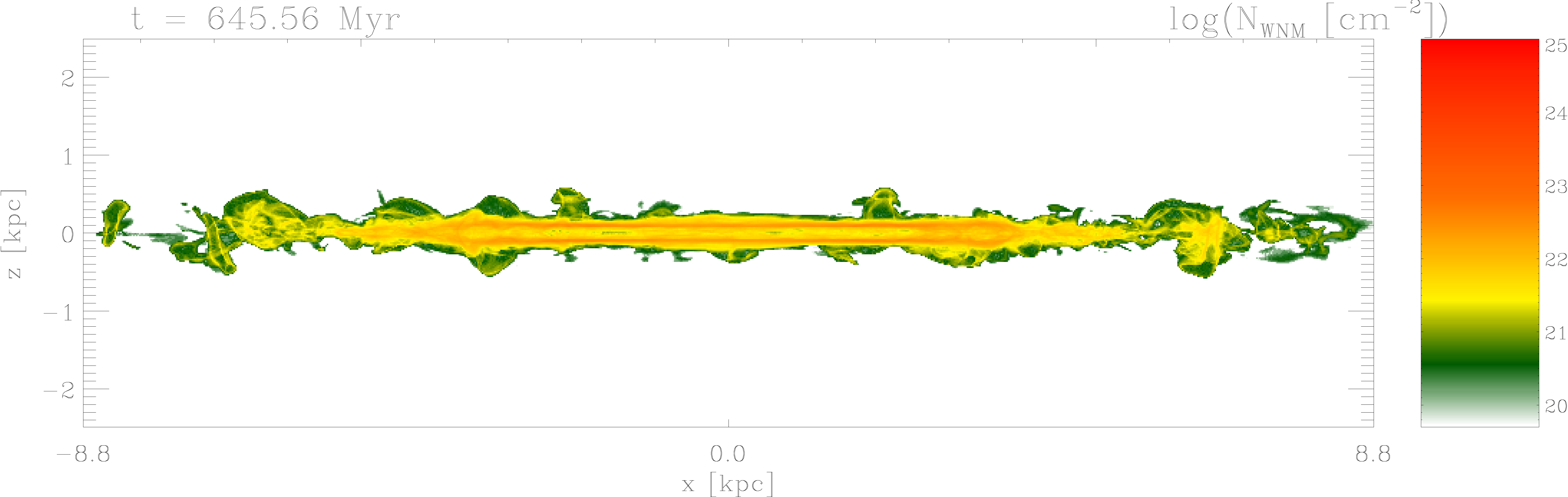}\\
  \rotatebox{90}{\qquad\qquad \fat{Beta0.25}}&\includegraphics[width=0.95\textwidth]{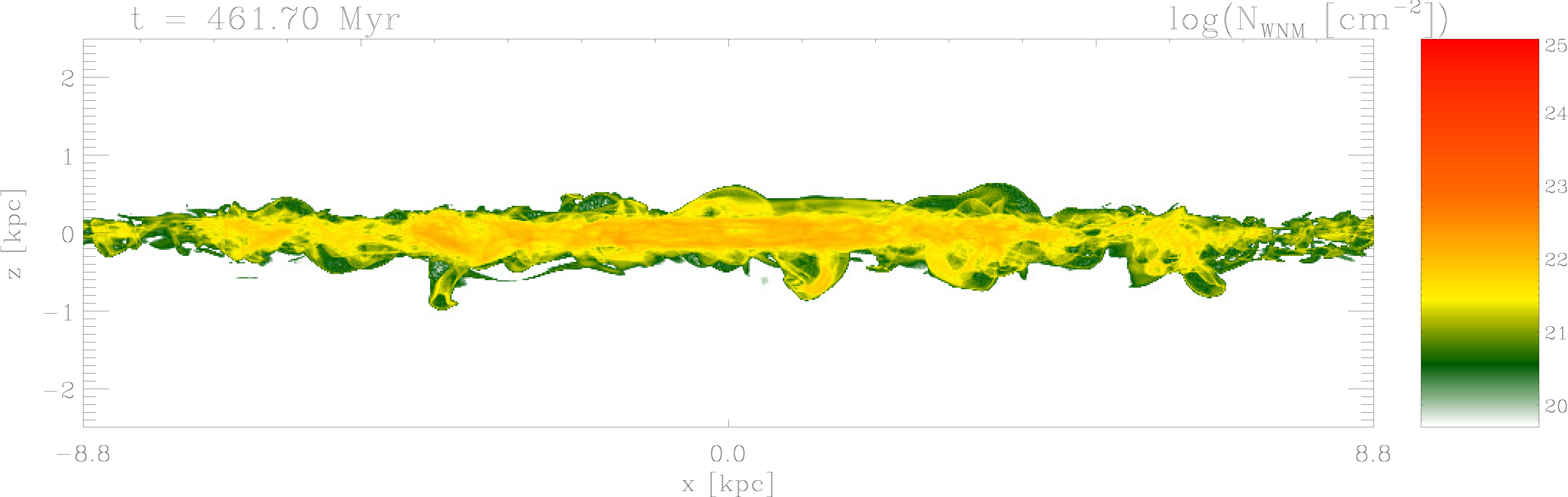}\\ \end{tabular}
\caption{Edge-on view of the WNM phase at times $t_\mathrm{onerot}$ for the three fiducial discs. Note the increasing disc thickness of all phases with increasing disc magnetisation. }
\label{figCdensEO}
\end{figure*}

\begin{figure*}
 \begin{tabular}{cc}
  \rotatebox{90}{\qquad\qquad \fat{Hydro}}&\includegraphics[width=0.95\textwidth]{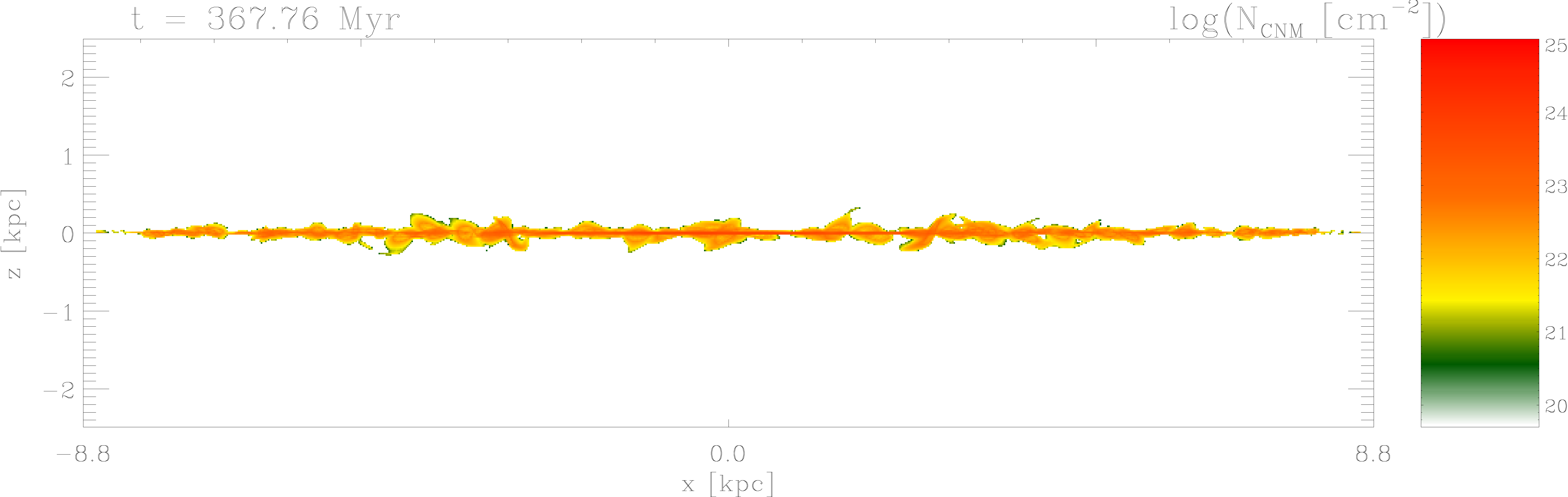}\\
  \rotatebox{90}{\qquad\qquad \fat{Beta1}} &\includegraphics[width=0.95\textwidth]{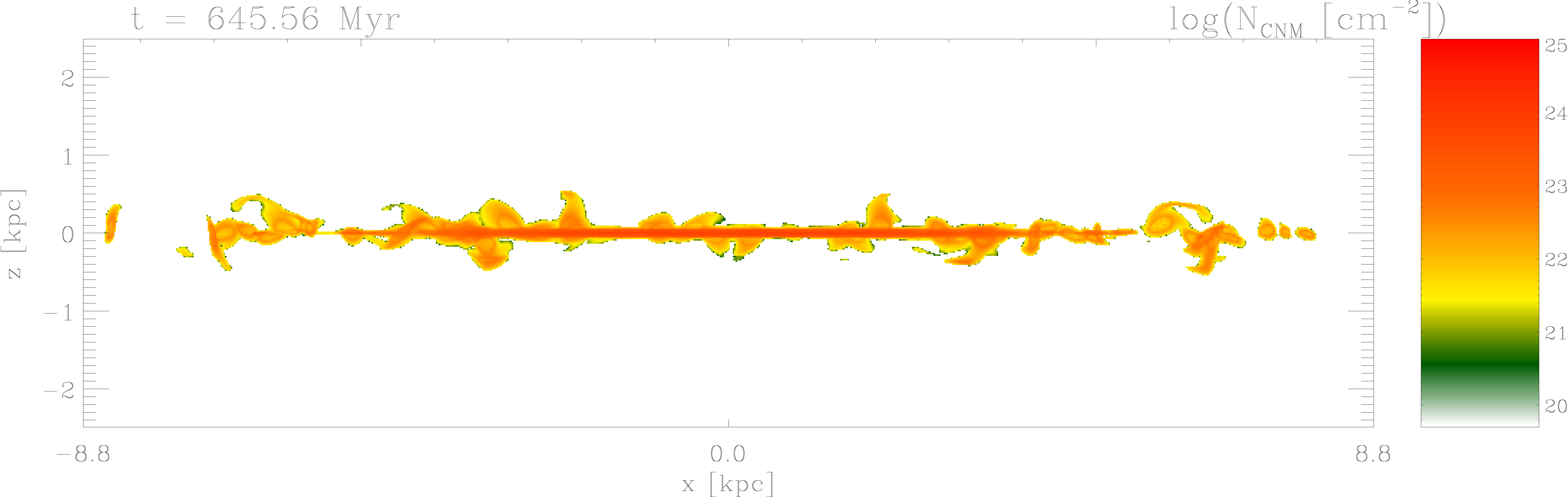}\\
  \rotatebox{90}{\qquad\qquad \fat{Beta0.25}}&\includegraphics[width=0.95\textwidth]{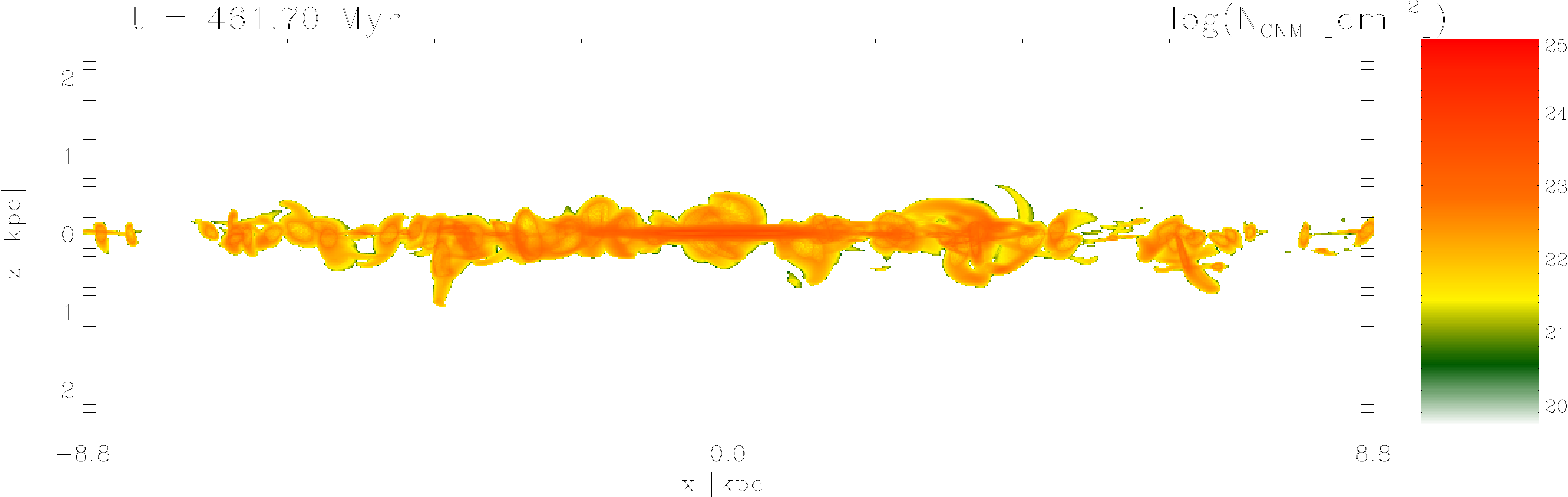}\\ \end{tabular}
\caption{Edge-on view of the CNM phase at times $t_\mathrm{onerot}$ for the three fiducial discs. }
\label{figCdensEO2}
\end{figure*}

\begin{figure*}
 \begin{tabular}{cc}
  \rotatebox{90}{\qquad\qquad \fat{Hydro}}&\includegraphics[width=0.95\textwidth]{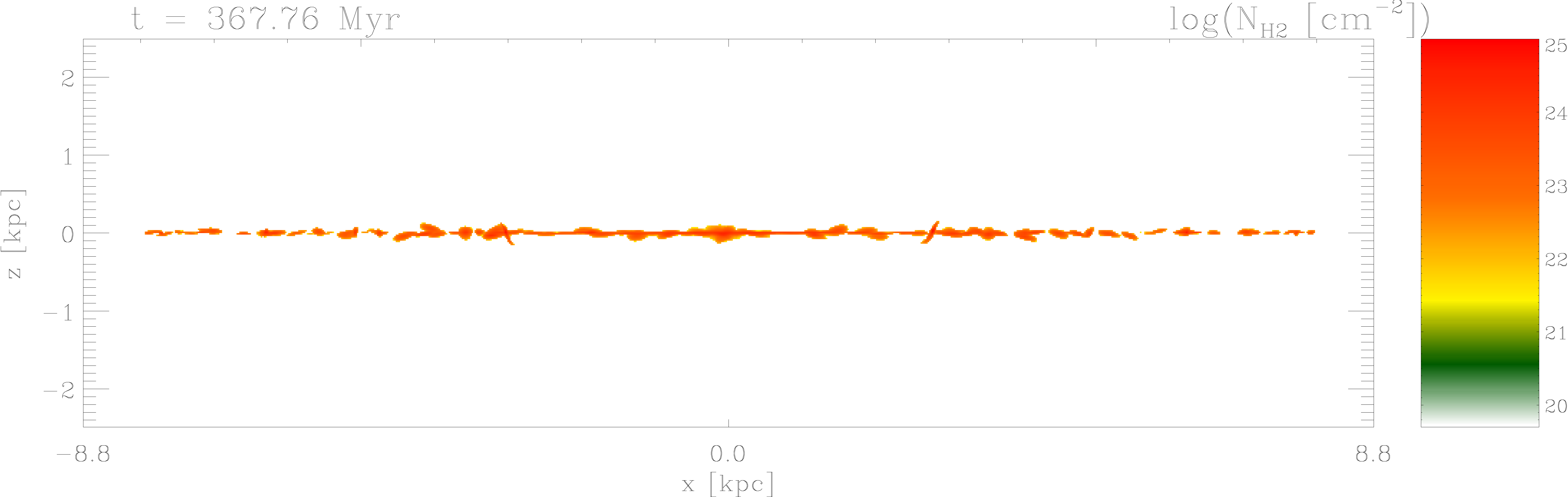}\\
  \rotatebox{90}{\qquad\qquad \fat{Beta1}} &\includegraphics[width=0.95\textwidth]{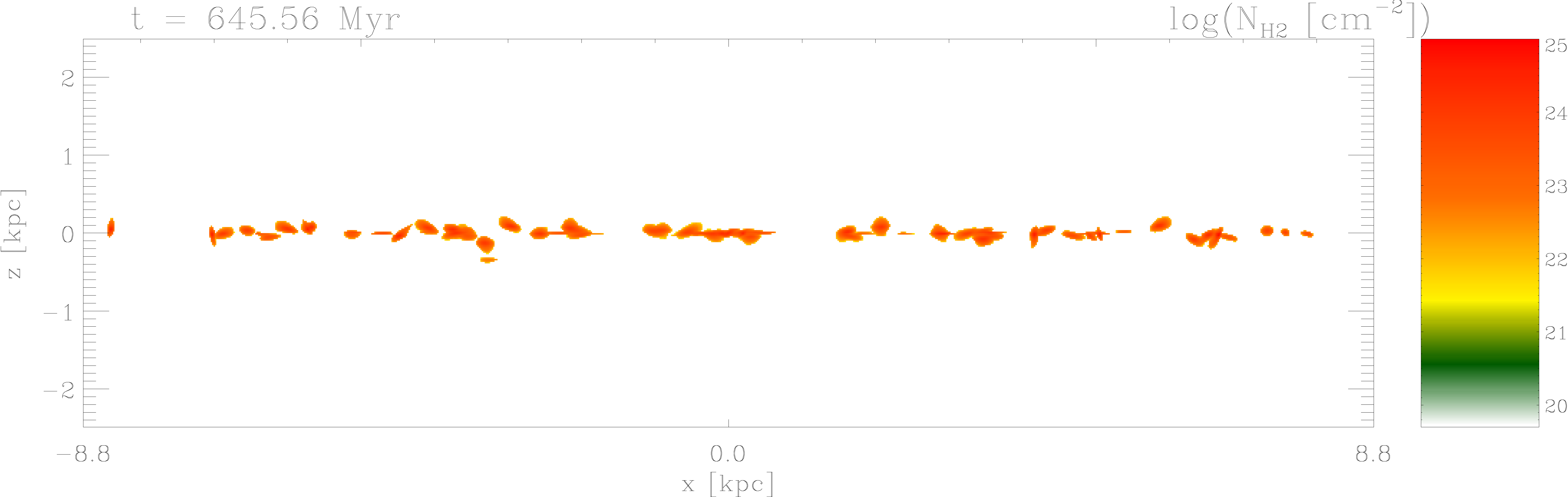}\\
  \rotatebox{90}{\qquad\qquad \fat{Beta0.25}}&\includegraphics[width=0.95\textwidth]{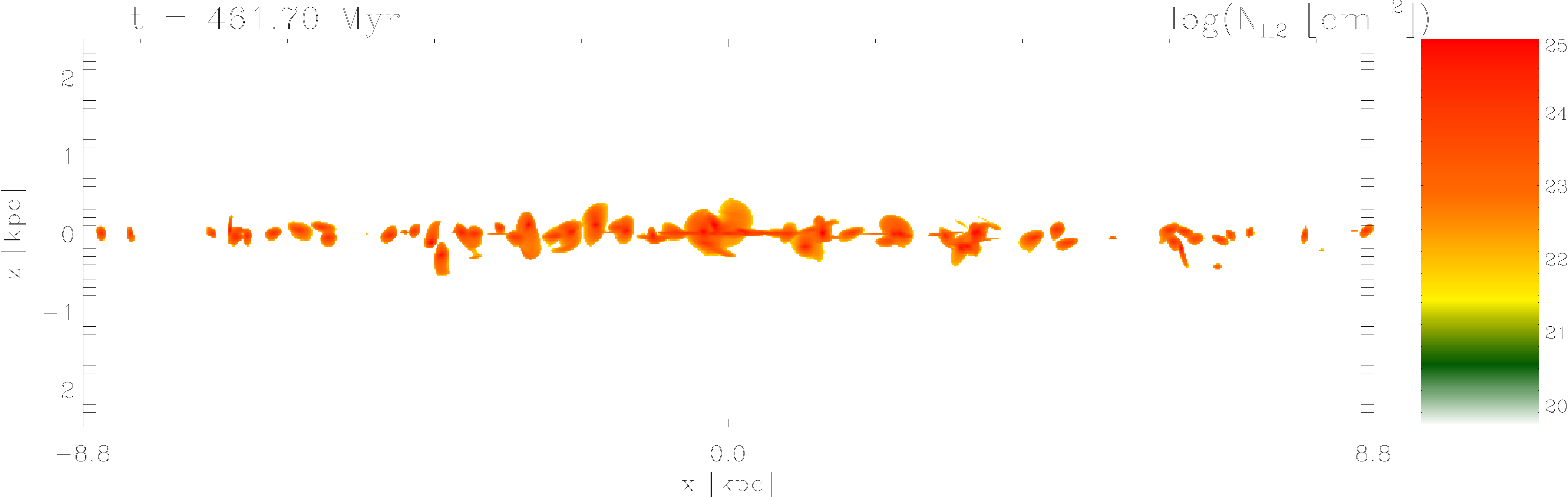}\\ \end{tabular}
\caption{Edge-on view of the CMM phase at times $t_\mathrm{onerot}$ for the three fiducial discs. Note the residence of cold, star-forming gas several hundred pc away from the midplane.}
\label{figCdensEO3}
\end{figure*}

\begin{table}
\centering
\caption{Range of gas temperatures used in this study to determine the properties of the galaxies.}
 \begin{tabular}{c|r}
    \hline
    \hline
    \fat{Name}	&\fat{Temperature Range}\\
    \hline
    WNM	&$1000<T/\mathrm{K}<5000$\\
    CNM	&$50<T/\mathrm{K}<300$\\
    CMM	&$20<T/\mathrm{K}<50$\\
    Star-forming (SF)&$T/\mathrm{K}<20$\\
    \hline
    \hline
 \end{tabular}
 \label{tabPhases}
\end{table}
The ISM is a multiphase gas, where different phases co-exist in approximate pressure equilibrium. For the remainder of this study, we separate the discs into various phases depending on their 
temperature. A list of the (four) chosen phases is given in Table~\ref{tabPhases}.\\
In Fig.~\ref{figCdensFO} we present face-on column density maps of the WNM, CNM and the cold molecular medium (CMM)-phase. In these 
maps, gas at temperatures not within the range specified for the respective phase does not contribute to the column density.
The discs reveal remarkable differences in the spatial distribution of the various gas phases\footnote{A discussion of the actual numerical values of the column density is not helpful due to the variations in 
the initial density profile.}. \\
The WNM distribution becomes increasingly smooth with increasing magnetisation of the disc. In the hydrodynamic disc, the WNM appears in shell-like structures, except 
for regions, where molecular clouds interact (see also Fig.~\ref{figColDensFO}). Here, the distribution is more extended. Using a finite plasma-$\beta$ instead reveals the emergence of WNM structures, which 
resemble filaments rather than shells. These filaments show lengths of several kpc and some are a few hundred pc thick. 
This morphological difference is, however, not a consequence of the fragmentation, but is rather related to the 
magnetic support of the gas against compression/disruption (in combination with the fact that there is no 
non-equilibrium heating/cooling. That is, gas, which is out of thermal equilibrium is immediately cooled/heated to the 
equilibrium temperature once the cooling timescale is smaller than the numerical timestep). Near over-dense regions, which are identified in the other gas phases (see 
below), the WNM distribution becomes more spherical. In the case of the strongest magnetic field, these spherical regions are entirely embedded in filamentary WNM structures, since the thickness of the 
latter becomes significantly large. We further acknowledge the larger number of filaments in the disc with $\beta=0.25$.\\
The CNM in the discs shows more similar morphological features. Most of the structures appear spherical, while a filamentary geometry becomes rarer. The remaining filaments correlate well with their 
WNM counterparts, i.e. CNM filaments only appear within the most extended (and possibly densest) WNM filaments. We emphasise that the CNM structures are not entirely smooth patches, especially in the 
hydrodynamic disc, where the CNM structures often appear as shells or as patches with varying column density.  In the magnetised discs no shells are seen. This 
increased smoothness of the CNM patches in the MHD discs suggests a less efficient mixing of the gas phases, though we caution that projection effects do play a role either.\\
On even smaller scales, corresponding to the CMM phase, both 
elongated and spherical objects are seen in the discs. These scales are primarily dominated by local dynamics. Differences in the column density of these objects are marginal and temporal effects.\\
We present edge-on views of the discs at $t_\mathrm{onerot}$ in Figs.~\ref{figCdensEO} to \ref{figCdensEO3}. As already discussed above, the magnetic field increases the vertical extent of the discs and hence also of the 
various gas phases. The WNM gas (Fig.~\ref{figCdensEO}) in the hydrodynamic disc extents up to only $\sim200\,\mathrm{pc}$ in regions, where clouds interact. The additional magnetic pressure enables the WNM to reside at far 
higher latitudes. The typical extent appears to be around $600-700\,\mathrm{pc}$ from the midplane, but some regions can be as far as $\sim1\,\mathrm{kpc}$. A similar trend is seen for the CNM gas, which 
is shown in Fig.~\ref{figCdensEO2}.
Especially noteworthy are the arc-like features above/below the midplane, which become increasingly pronounced in the magnetised discs due to gas flows along curved field lines (see also Fig.~\ref{figColDensEO}). Interestingly, the 
gas away from the disc midplane becomes sufficiently dense that it undergoes a 'phase-transition' to the CMM-phase, as indicated in Fig.~\ref{figCdensEO3}. This latter fact 
promotes not only the 'general' levitation of material out of the midplane, but also the residence of dense and molecular structures at such high latitudes.

\subsection{Disc dynamics}
In the following, we discuss the time evolution of selected quantities of all simulated galaxies. As above, we split the gas into four phases according to their temperature. The focus will lie on the 
gas within $|z|\leq300\,\mathrm{pc}$ and \mbox{$0.5<R/\mathrm{kpc}\leq10$}, that is, on the main disc. We emphasise that quantities, which are related to the turbulent velocity field, are not converged due to the
 limited spatial resolution. As also discussed in \citet{Jin17} and \citet{Koertgen17b} the spatial resolution should be $\lesssim0.1\,\mathrm{pc}$ for the velocity field to converge.
\subsubsection{Mass}
\begin{figure*}
\centering
\includegraphics[width=0.5\textwidth,angle=-90]{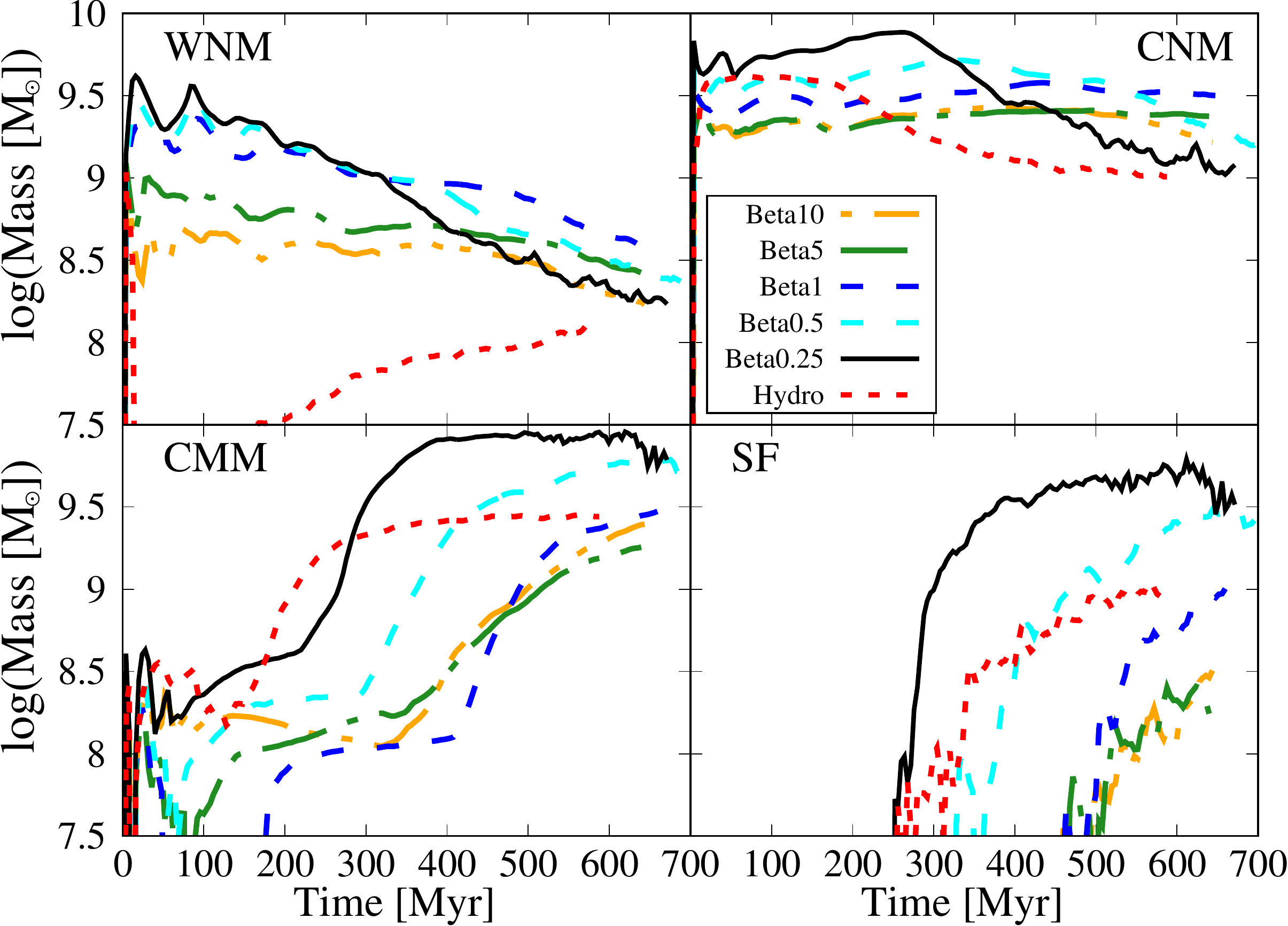}
\caption{Time evolution of the mass in the different gas phases within $|z|\leq300\,\mathrm{pc}$. We point out that differences due to the initial disc magnetisation become pronounced only in the 
coldest phases and primarily as a temporal delay.}
\label{figTimeMass}
\end{figure*}

The time evolution of the mass in the four phases is shown in Fig.~\ref{figTimeMass}. The phases of the magnetised discs evolve in a similar way, though variations can appear at different times. At very early 
times, a sharp decrease is seen. This is due to the fact that the disc is initially not in hydrostatic equilibrium and thus has to evolve to such a state. Furthermore, all galaxies get 
compressed due to the action of the (strong) external potential. This compression, if strong enough, induces a phase transition to a different phase and thus reduces the amount of gas (hence, mass) in the WNM. Later, the mass in 
the WNM does not show a significant evolution up to $t\sim300-400\,\mathrm{Myr}$. From this time on, the WNM mass decreases in all MHD discs, eventually converging on quite similar values. In contrast, the 
hydrodynamic disc reveals a significant drop in the total mass of the WNM at early times up to $t\sim70-90\,\mathrm{Myr}$. This suggests that the (toroidal) magnetic field plays a stabilising role for the WNM. 
After this sharp drop, the mass in the 
WNM starts to increase again relatively quickly and saturates at a few $10^9\,\mathrm{M}_\odot$.\\
We next discuss the CNM with temperatures \mbox{$50\leq T/\mathrm{K}\leq300$}. The discs with $\beta>0.25$ show no pronounced time variations in the CNM mass. Only the two extreme cases with $\beta=0.25$ and 
$\beta=\infty$ reveal some variation. The mass in the CNM of the hydrodynamic disc increases at early times $t\lesssim50\,\mathrm{Myr}$, stays almost constant up to $t\sim200\,\mathrm{Myr}$ and decreases 
afterwards. The initial increase is a result of the disc compression and the subsequent phase transition of warm to cold gas. This is followed by a phase of no evolution in the CNM, most likely due to the 
disc reaching some quasi-equilibrium state. At later times, fragmentation of the disc induces a decrease of the mass in the CNM due to further phase transitions both towards colder and warmer phases.\\
Contrary to the evolution of the hydrodynamic disc, the mass in the CNM of the strongly magnetised disc increases at first. The maximum mass in the CNM is reached between $t\sim250-300\,\mathrm{Myr}$ and 
decreases afterwards due to violent fragmentation of the disc. In the end, the hydrodynamic and magnetised discs reveal comparable amounts of mass in the CNM.\\
As star formation is strongly associated with cold, dense and molecular gas, we next focus on the CMM-phase before discussing the actual star-forming gas. The total mass is observed to be comparable during the very early stages of evolution, though some temporal fluctuations occur. In some cases, 
the total mass in this gas phase even drops below $10^8\,\mathrm{M}_\odot$ for some time. However, overall, the evolution is very similar. The initial phase with some variation is followed by a sharp 
increase in the mass of this phase. The increase in mass is about an order of magnitude and this stage lasts for about 100\,Myr. The hydrodynamic disc increases its mass first, followed by the disc with 
initial plasma-$\beta=0.25$. The late phase of disc evolution is characterised by a constant total mass in the CMM-phase, where the masses of the more strongly magnetised discs become comparable.\\
Considerable amounts of mass in the star-forming phase are generated from $t\sim300\,\mathrm{Myr}$ on. This phase has typical densities of $n\sim10^4\,\mathrm{cm}^{-3}$. Interestingly, the disc with 
$\beta=0.25$ first generates a large reservoir of mass in this phase. Hence, the efficiency to transfer gas from the CMM- to the SF-phase is larger in the magnetised discs, as further supported by the 
quick increase in mass for the disc with initial $\beta=0.5$. Since the mass approaches a plateau at late times, the effect of the magnetic field is to support the dense/cold gas against shear flows. Note 
that the mass in the SF-phase grows somewhat slower in the hydrodynamic disc, likely due to the lack of support against shearing motions.\\
The above note is closely related to the PI. The global MHD PI is able to generate filaments and flows extending over many kpc. This addresses Mestel's point that in order to form a GMC, the extensive gas supply needed  
must be gathered very efficiently from large pieces of the galaxy. The old PI instability picture addressed by \citet{Elmegreen82a,Elmegreen82b} pictures only a rather 2D picture with gas falling into a local magnetic 
minimum.  In our simulations, gas falls into kpc long magnetic valleys, and quickly gathers into the dense, cool phase.  This also addresses critical arguments to the PI role in cloud making made in 
\citet{McKee07}.  

\subsubsection{Mass fraction}
\begin{figure*}
	\includegraphics[width=0.5\textwidth,angle=-90]{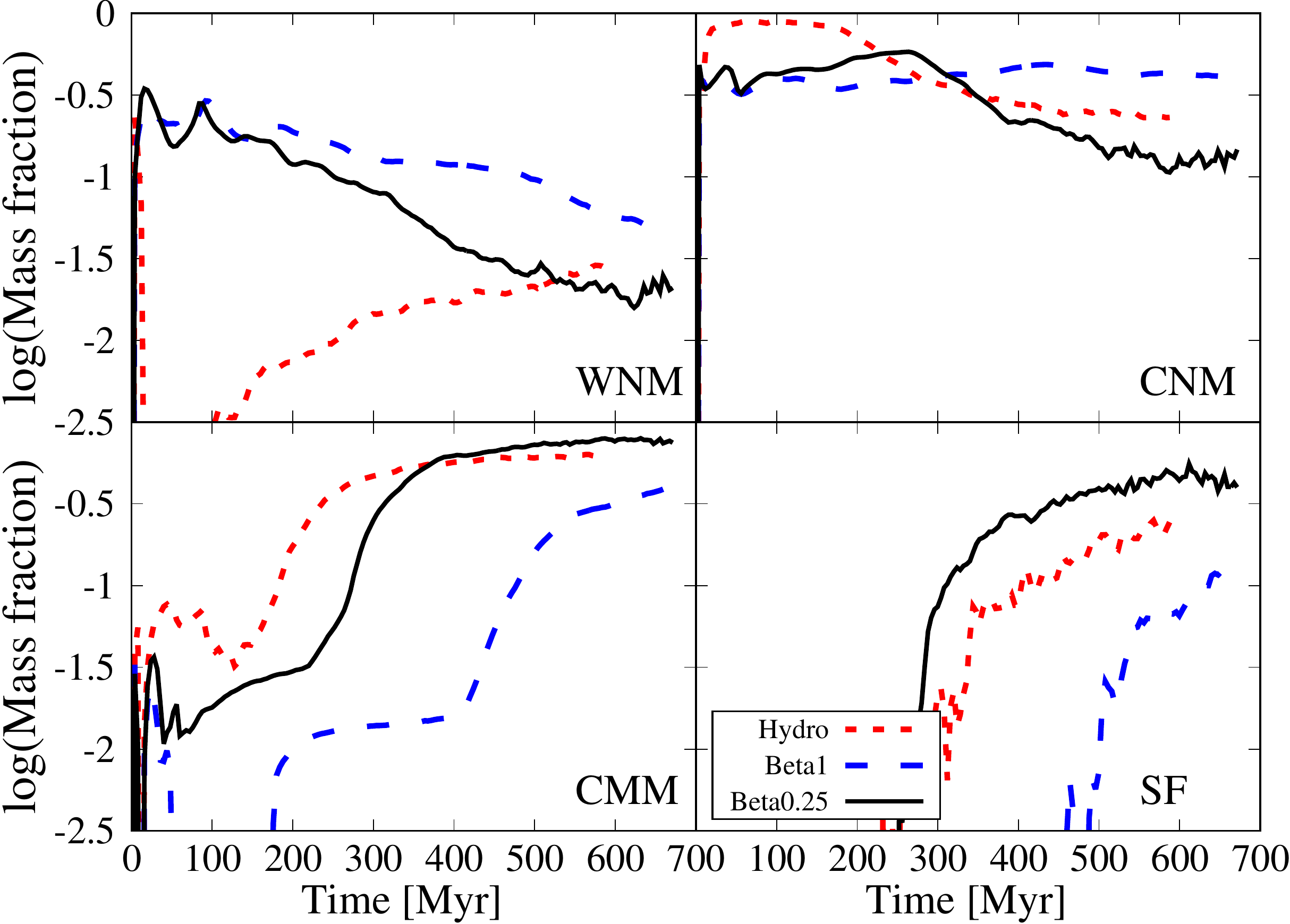}
	\caption{Time evolution of the mass fraction of the different phases within $\left|z\right|\leq300\,\mathrm{pc}$.}
	\label{figMassfraction}
\end{figure*}

To remove differences in the mass evolution due to variations in the initial density profile, we highlight the time evolution of the mass fraction in Fig.~\ref{figMassfraction} for the three fiducial 
galaxies (characterised by $\beta=\left\{\infty,1,0.25\right\}$). \\
The mass fraction of the WNM shows a clear difference between the magnetised and the hydrodynamic galaxies. While the mass fraction slowly decreases for the MHD cases, it sharply drops in the hydrodynamic 
disc. This drop is primarily associated with the expansion of the outer parts of the disc in the radial direction and the additional compression along its vertical axis. The compression induces a phase transition to the 
colder phase, while an expansion moves gas out of the analysis region. Interestingly, the MHD mass fractions do not differ by much in the first 150 Myr. From this time on, pronounced differences become 
visible, since the disc with $\beta=0.25$ fragments first, thus pushing gas towards other phases. The hydrodynamic disc reveals an increasing mass fraction of the WNM towards later times, with the fraction 
becoming comparable and even larger than in the strongest magnetised disc. Towards the end of each simulation, the mass fraction of the WNM in the three cases approaches 1-3\,\%.\\
The CNM fraction reveals less pronounced decreases. However, a clear distinction can be made between early and late fragmenting discs. While the disc with initial $\beta=1$ shows no clear variation due to 
a very late fragmentation, a 
decrease is seen for the discs with $\beta=\left\{\infty,0.25\right\}$ as gas is pushed from the CNM towards the even denser/colder phases in the absence of any form of (stellar) feedback. At the end of the 
simulation, about 10-30\,\% of the gas reside in this phase.\\
Without stellar feedback, gas is continuously transferred to a colder phase. This is readily seen in the CMM-phase, where the mass fraction of this phase sharply rises once the discs have begun to 
fragment into small-scale clouds. The relative increase is around an order of magnitude in all discs, though the onset of the increase depends on the initial conditions. In the end $\gtrsim30\,\%$ of the gas 
reside in this phase. The strongest magnetised disc shows a slightly larger mass fraction at late times. The disc with $\beta=1$ is still in a phase of increasing mass fraction. On comparing these data with the 
ones from the CNM phase, it becomes clear that the CNM phase is replenished much quicker by gas from warmer phases, since the drop of the mass fraction is much less pronounced as the rise in the 
CMM-phase. \\
Last, we study the evolution of the star-forming gas phase with $T\leq20\,\mathrm{K}$. These temperatures typically correspond to gas densities $n\gtrsim10^4\,\mathrm{cm}^{-3}$. Here, again, the 
outlier is the disc with initial $\beta=1$ due to the far later onset of disc fragmentation. Similarly large mass fractions in this phase are obtained at around $t\sim250\,\mathrm{Myr}$ for the hydrodynamic 
and the strongest magnetised disc. However, with time, the evolution of these discs diverges. The hydrodynamic evolution is observed to be much more varying, while its MHD counterpart is smoother. 
In the end, the MHD disc reaches mass fractions of $>30\,\%$ for the star-forming phase, slightly larger than the hydrodynamic disc. However, we note that the latter disc shows that the mass fraction in the 
star-forming phase is still increasing.\\
Generally, this significantly different evolution in the mass fraction of the very cold (thus dense) gas - the CMM and star-forming phases - implies that the magnetic field has two effects. The first is that these 
phases do not undergo strong variations in their mass fractions. This reveals that the field supports the gas against disruption by shearing motions. The second, and most likely more important, effect is 
that the flow of gas from the CMM towards the star-forming phase is increased / much more efficiently in the magnetised discs. This can be inferred by comparing the mass fractions of the two phases 
at certain times. At around $t\sim250\,\mathrm{Myr}$, the mass fraction of the CMM-phase for the MHD disc is smaller by about an order of magnitude, while the fraction of gas in the star-forming 
phase is similar. At around $t\sim380\,\mathrm{Myr}$, the fractions of gas in the CMM-phase are comparable, but the amount of gas in the star-forming phase is much higher for the magnetised disc. \\
To sum up, the effect of the field on the gas evolution appears to be small on large scales (represented by the WNM and CNM) towards late times, but is observed to be larger on smaller scales (that is, in the denser 
gas), i.e. by inducing a faster conversion towards the coldest phase in the disc.

\subsubsection{Ratio of thermal to magnetic pressure}
\begin{figure*}
\centering
\includegraphics[width=0.5\textwidth,angle=-90]{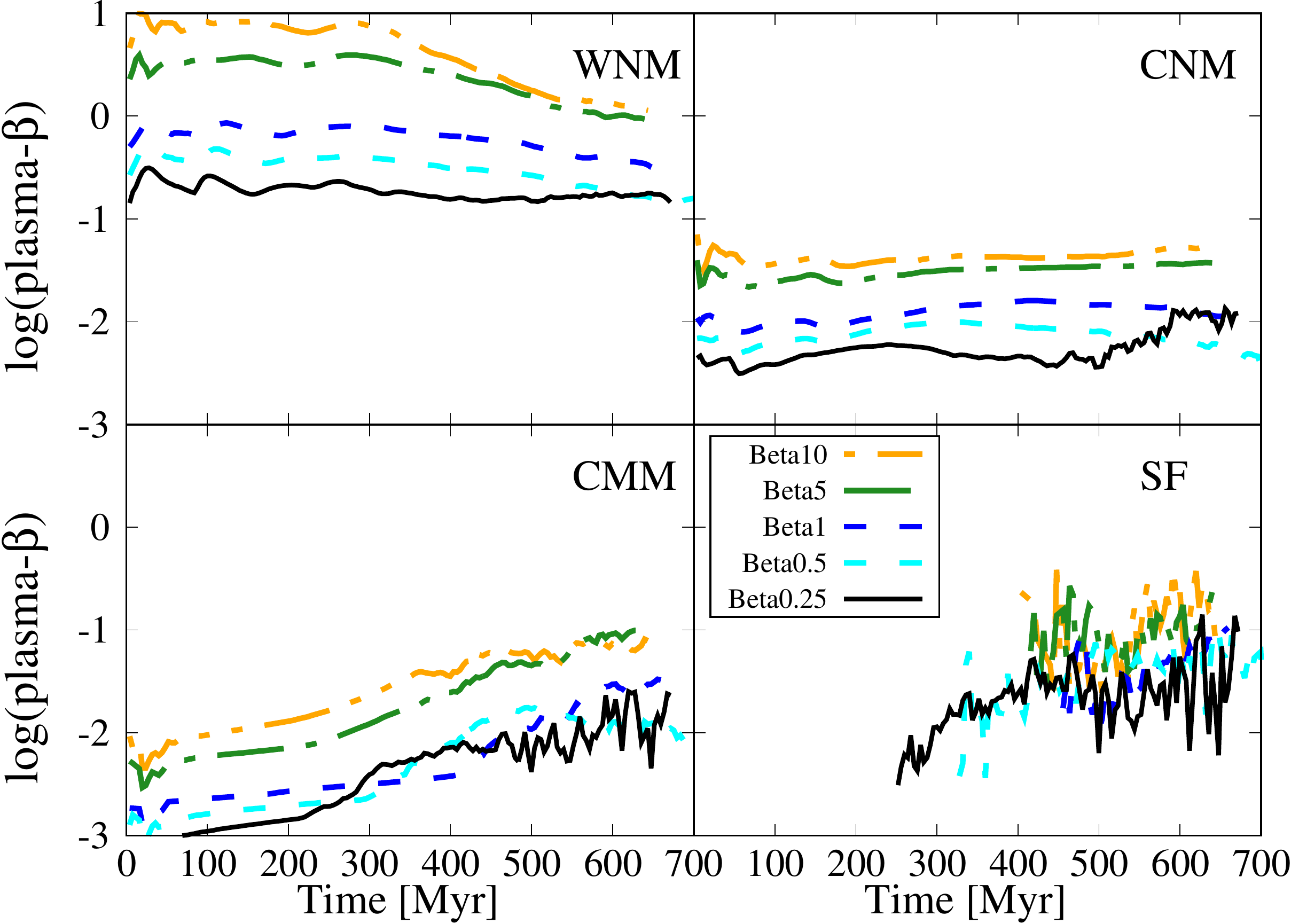}
\caption{Time evolution of the plasma-$\beta$ in the different gas phases within $|z|\leq300\,\mathrm{pc}$. Note the generally different evolution of $\beta$ in the WNM/CNM and CMM/SF phases.}
\label{figBeta}
\end{figure*}

The evolution of $\beta$ for all magnetised discs is given in Fig.~\ref{figBeta}. 
Interestingly, the time evolution of this quantity clearly depends on the phase. While the warmer phases - here the WNM and CNM - show no or only little temporal variations, the cold phases (the 
CMM and star-forming) reveal fluctuations, which can be as large as an order of magnitude. Besides the lack of temporal variations, the $\beta$-values of the WNM and CNM stay close to their 
initial value for a long time. This is especially prominent in the disc with $\beta=0.25$. Here, the smallest variations are seen, while for the other discs, a slight decrease is seen towards later times.
It is observed that the discs with initial $\beta>1$ reach equipartition between thermal and magnetic energy towards the end of the simulation. The discs with $\beta\leq1$ decrease towards 
$\beta\sim0.1-0.2$. From the decreasing trend of the latter three discs, we expect the two discs with initial $\beta>1$ to evolve to a state in which the magnetic field represents the dominant form of 
pressure support in the WNM, consistent with extragalactic observations \citep{Beck15,Beck16}. The CNM in the discs shows less evolution and is even more dominated by magnetic fields, as expected since 
the gas is cooling down. In contrast to the WNM, the ratio of thermal to magnetic pressure reveals no obvious trends.\\
The colder phases in the discs reveal a more pronounced time evolution. The general trend is that $\beta$ increases over time in these phases, but seems to saturate at least within the star-forming phase. 
There might even be a saturation for the discs with $\beta=\left\{0.25,0.5\right\}$ at late stages in the CMM-phase, but this is less obvious. The reason for $\beta$ becoming larger is that the discs undergo 
fragmentation via the Parker instability. Material is compressed by flows along the magnetic field lines. Magnetic flux is not dragged along into these regions, hence only the density increases. If the 
density increase is faster than the temperature \ita{decrease}, thermal pressure will rise and so will the plasma-$\beta$. Interestingly, the final values achieved are of the same order as those of the CNM and 
WNM. Similar values are obtained in the colder phase of the star-forming gas, but with much more temporal fluctuations, which might be caused by the artificial pressure floor used in our calculations.

\subsection{Magnetic levitation and high latitude star-forming gas}	
\begin{figure*}
	\includegraphics[width=0.5\textwidth,angle=-90]{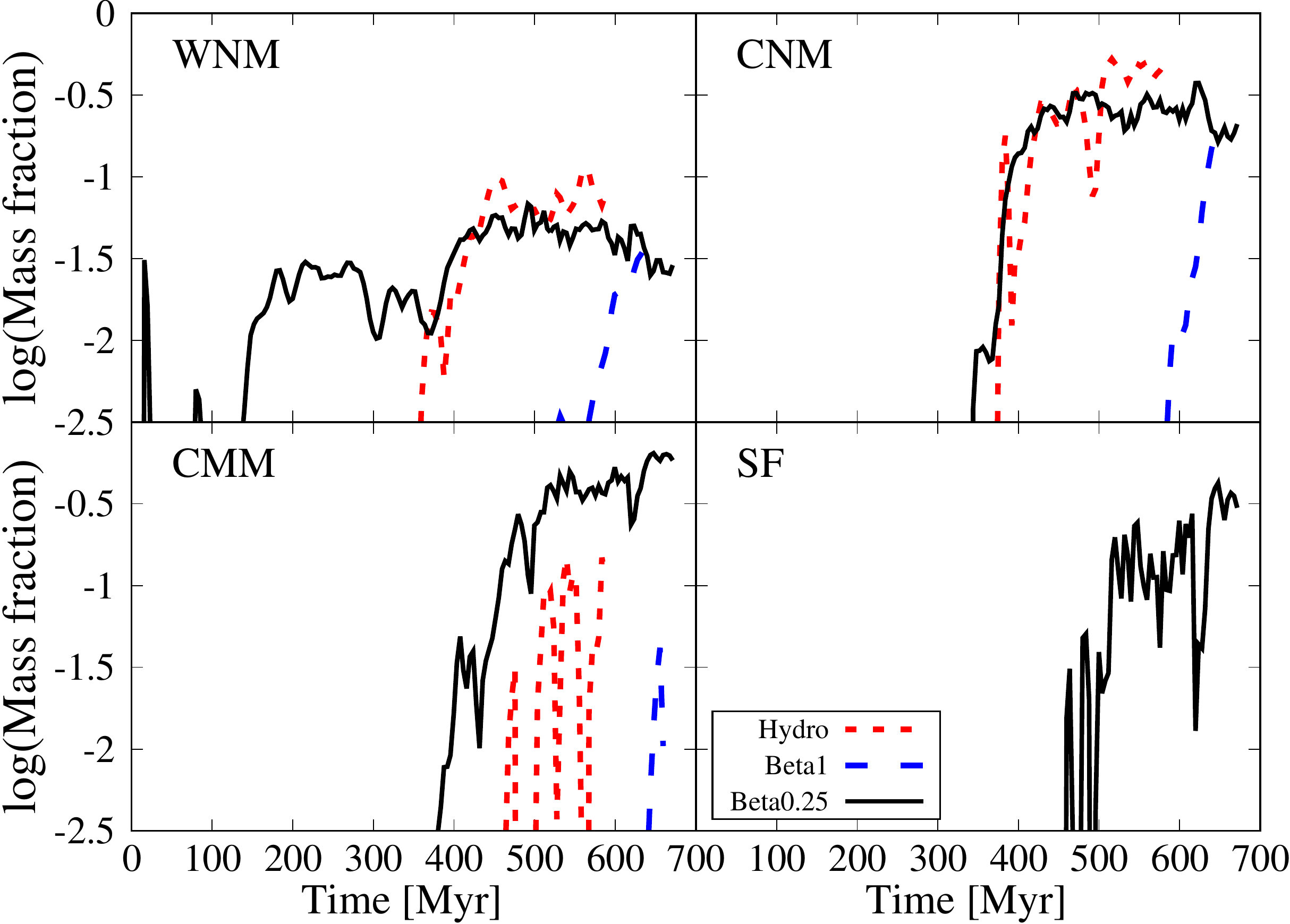}\\
	\caption{Time evolution of the mass fraction at high latitudes $300\,\mathrm{pc}\leq\left|z\right|\leq2000\,\mathrm{pc}$ for the same discs as in Fig.~\ref{figMassfraction}.}
	\label{figMassfraction2}
\end{figure*}

\begin{figure*}
 \begin{tabular}{ccc}
    \includegraphics[width=0.31\textwidth,angle=-90]{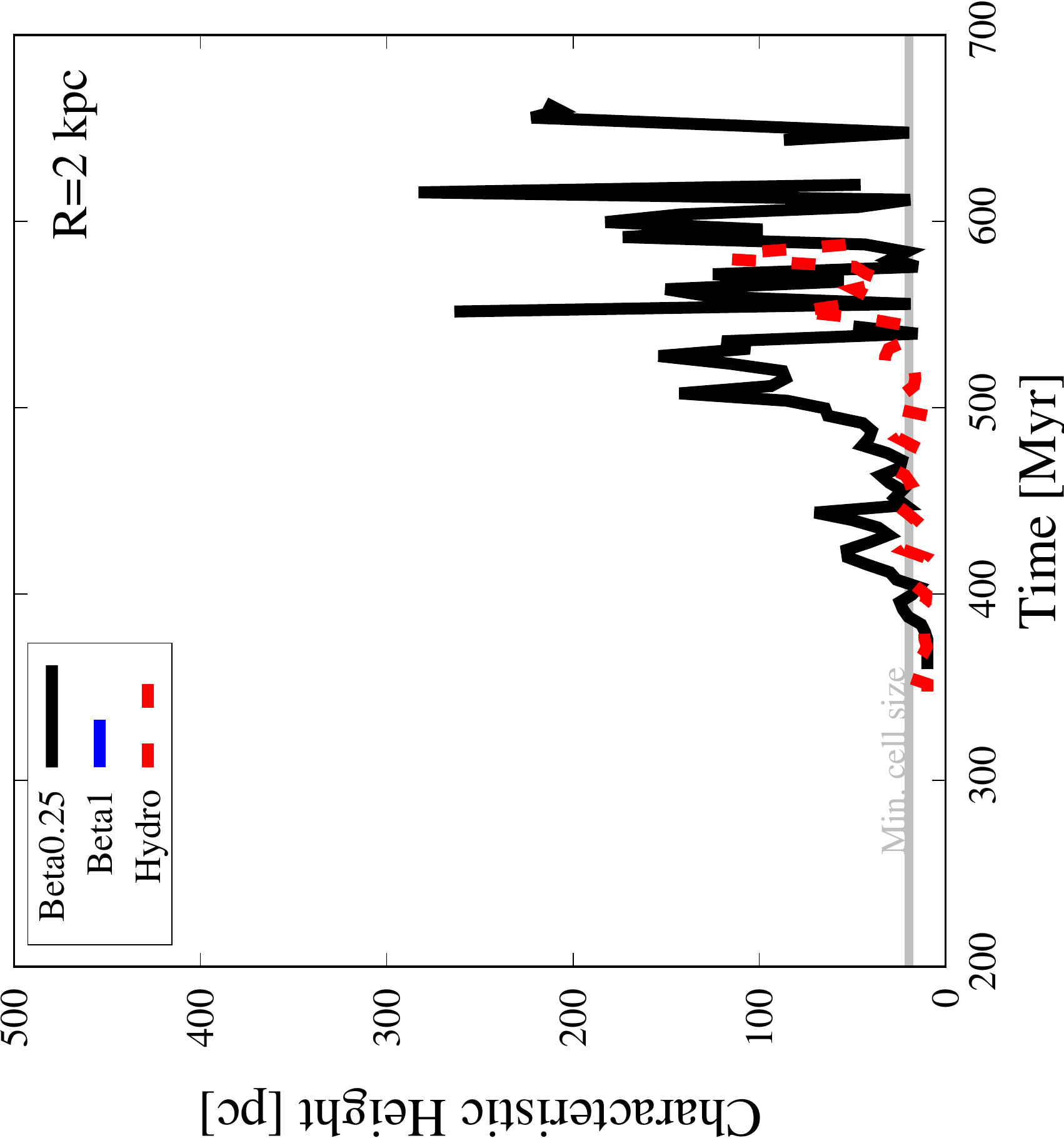}
    &\includegraphics[width=0.31\textwidth,angle=-90]{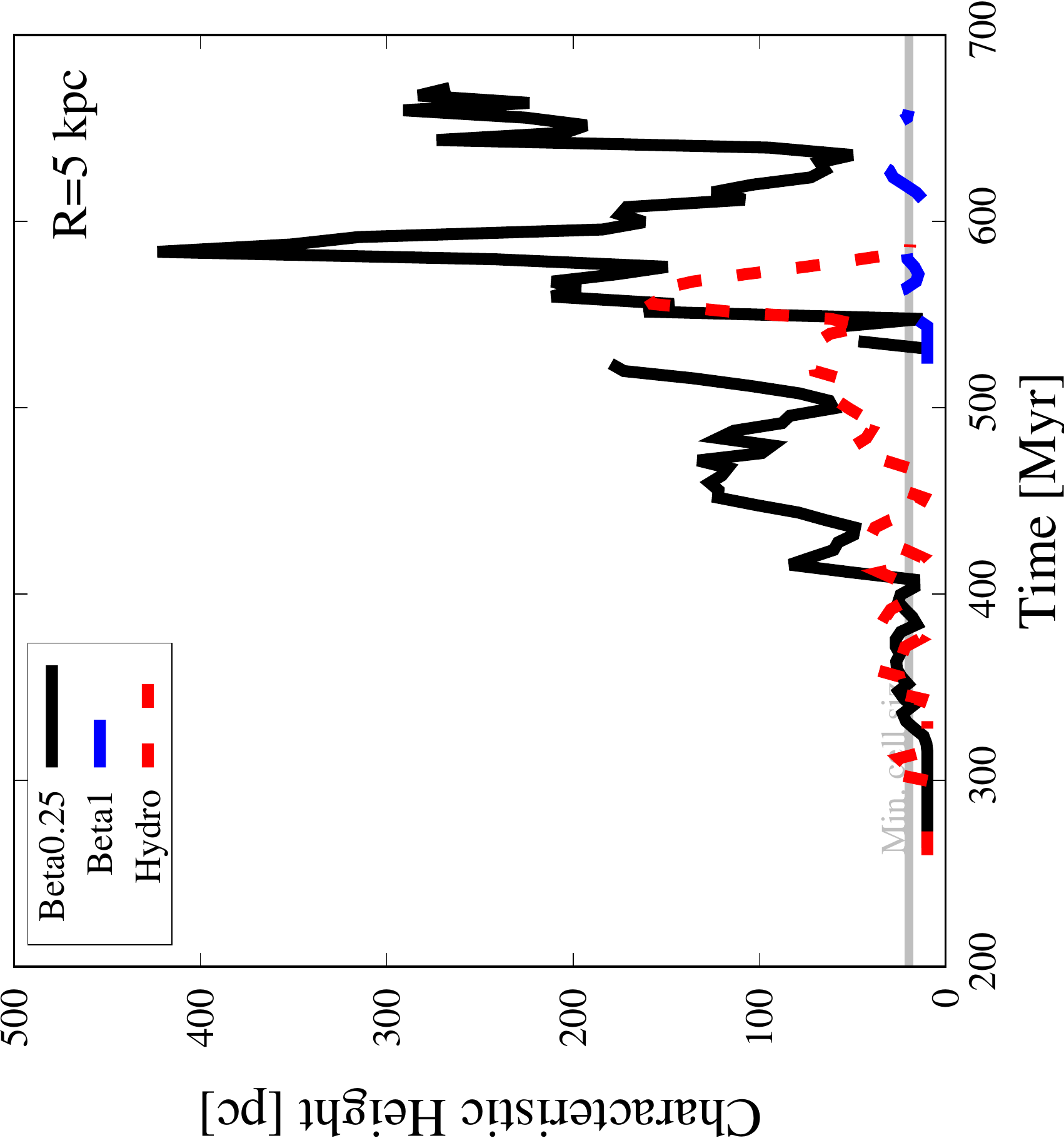}
    &\includegraphics[width=0.31\textwidth,angle=-90]{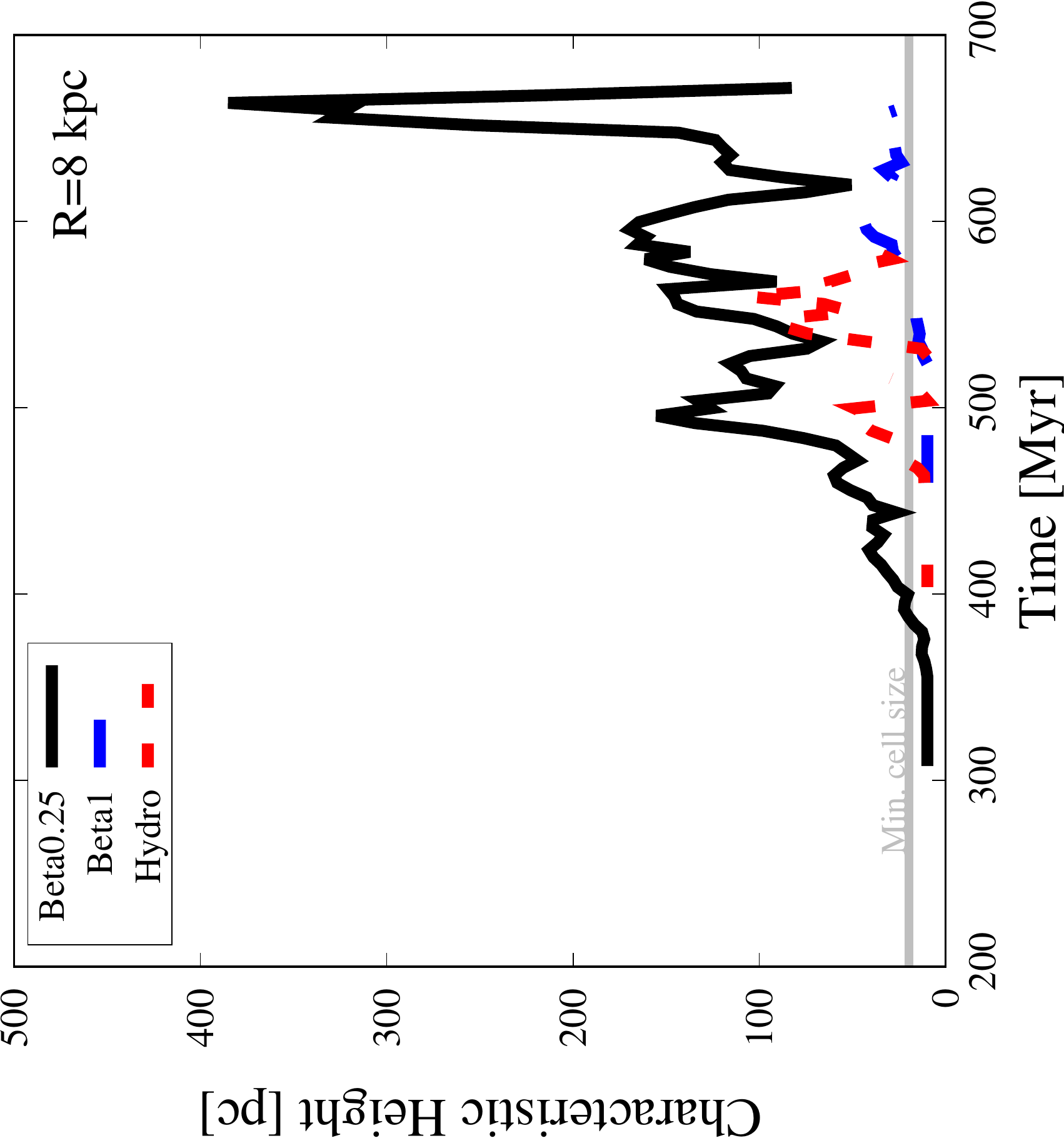}\\
     \includegraphics[width=0.31\textwidth,angle=-90]{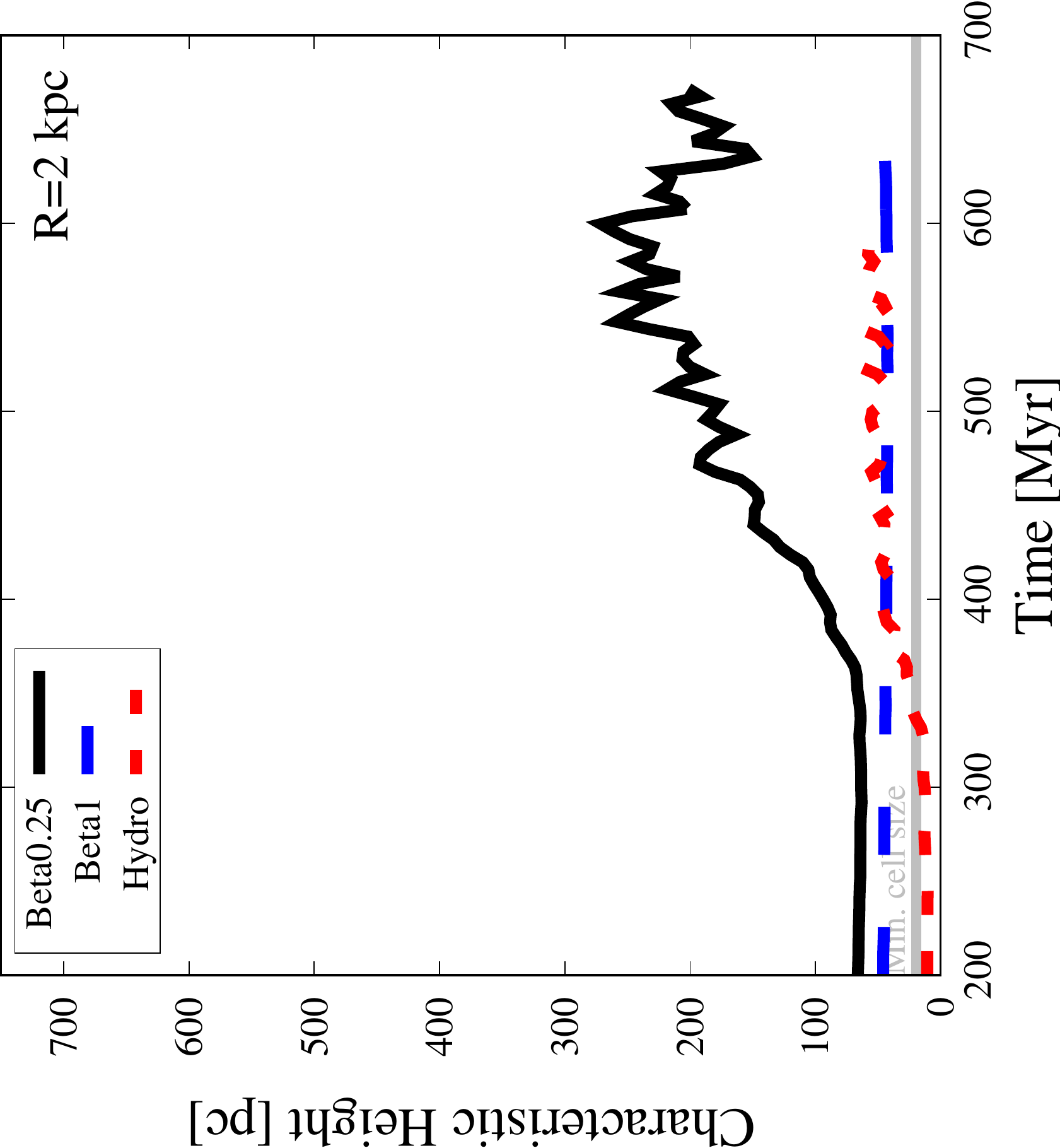}
    &\includegraphics[width=0.31\textwidth,angle=-90]{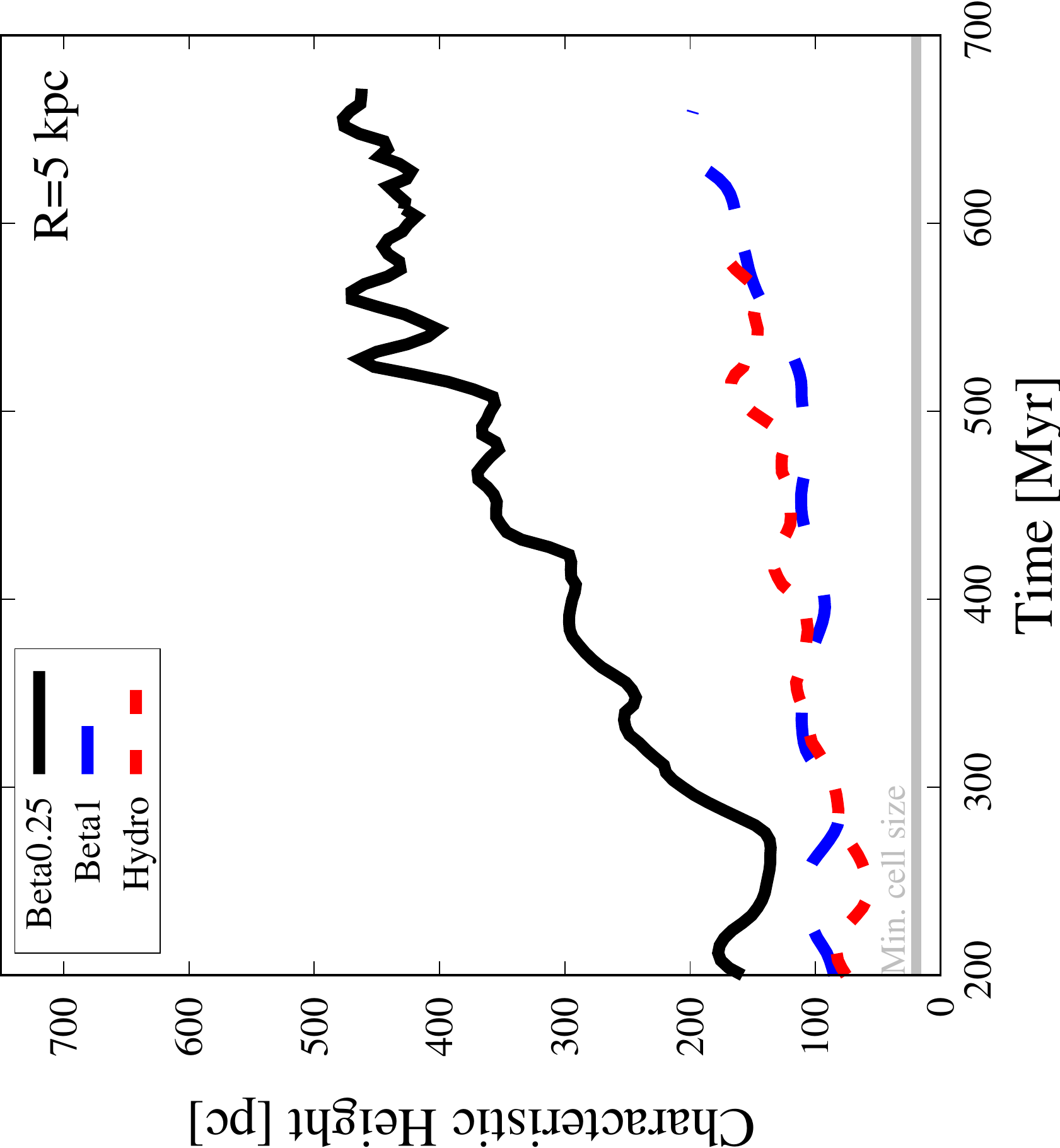}
    &\includegraphics[width=0.31\textwidth,angle=-90]{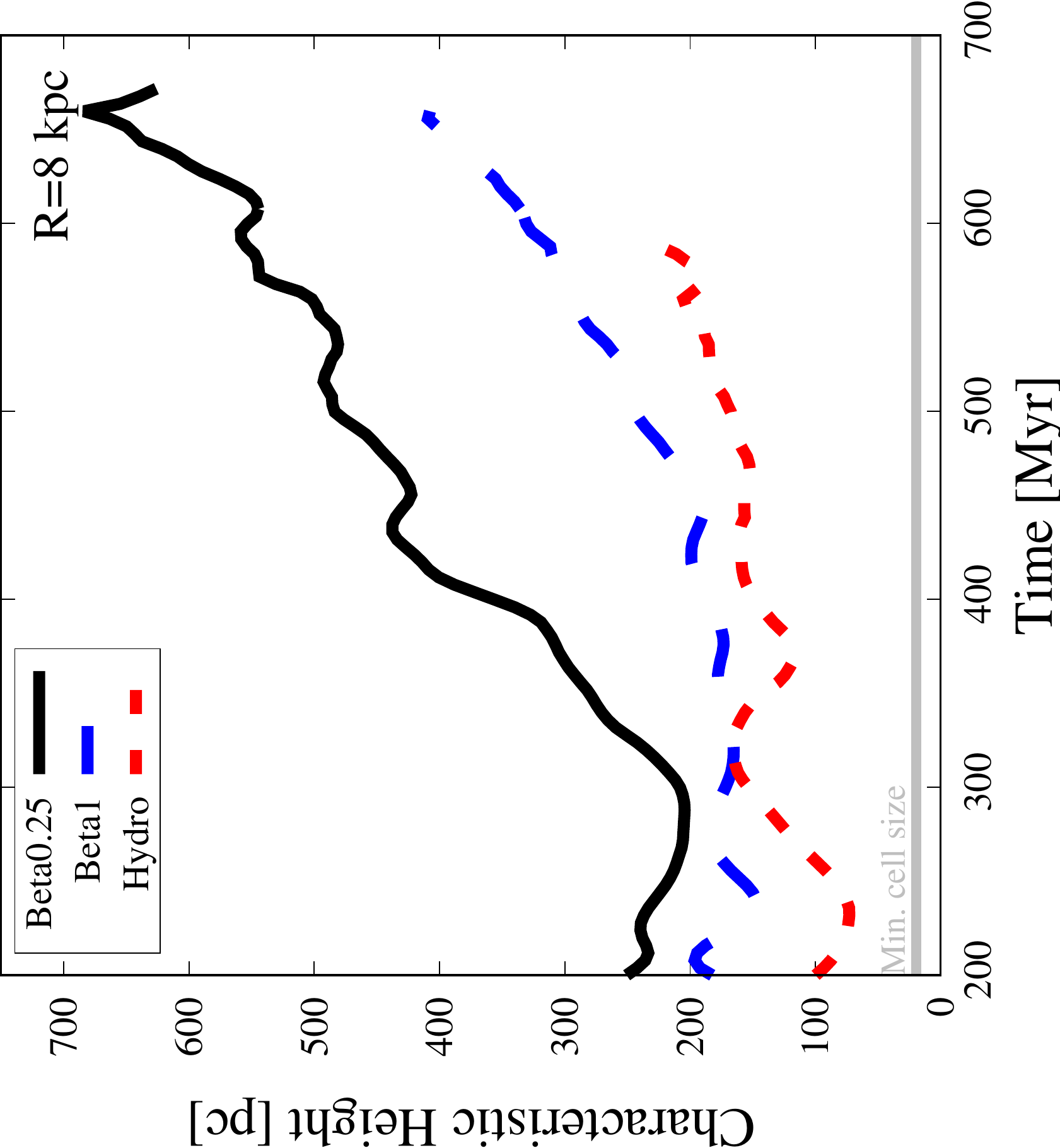}
 \end{tabular}
 \caption{The characteristic height, $H_\mathrm{char}$ as a function of time for various annuli. Top row depicts star forming gas with $T\leq20\,\mathrm{K}$, bottom row for 
 the CMM gas with $20<T/\mathrm{K}\leq50$.}
 \label{figCharHeight}
\end{figure*}

In Fig.~\ref{figMassfraction2} we present the time evolution of the mass fraction of the four phases for heights above/below the disc midplane of $300\,\mathrm{pc}\leq\left|z\right|\leq2000\,\mathrm{pc}$. 
In general, all but the star-forming phase are represented at such high galactic latitudes for \ita{all} discs. While the WNM phase appears at these latitudes rather early for the strongest magnetised galaxy due to disc thickening because of the 
additional pressure component, 
a significant fraction in the hydrodynamic or $\beta=1$-scenario only appears at late stages $t>350\,\mathrm{Myr}$. For the $\beta=1$ disc this is reasonable, as it fragments rather late. In the hydrodynamic 
case, the Toomre instability acts preferentially along the radial direction. Hence, all subsequent dynamics are at first confined to the galactic midplane. Only later, when local dynamics - e.g. via cloud-cloud 
interactions - become dynamically important, an efficient transport of material along the vertical direction is initiated. However, in all three shown discs, the mass fractions saturate at around a few percent.\\
The magnitudes of the mass fraction of the CNM at these heights are comparable for the galaxies at late times. Interestingly, the hydrodynamic and the strongest magnetised disc show a similar evolution, 
while the disc with $\beta=1$ is again delayed due to the later fragmentation. The similarity in the CNM evolution here is due to the previously mentioned dynamical argument of cloud-cloud interactions 
pushing cold material to higher latitudes. These processes appear in all discs, independent of their magnetisation. However, studying the CMM-phase, it becomes evident that the material, which is being 
lifted upwards, is not as cold (or dense) in the hydrodynamic disc. The mass fraction of the CMM gas is much larger than the one in the hydrodynamic disc. This indicates that cloud-cloud interactions 
cannot be the only cause for such cold material at these heights. These interactions should account for temporary peaks in the mass fractions, as is truly seen in the hydrodynamic case, as collisions or 
mergers of clouds are short-period scenarios.\\
The continuous increase of the mass fraction for the disc with $\beta=0.25$ points to a magnetic origin. However, spiky features are identified over time, which 
are indeed due to the previously mentioned cloud-cloud collisions. 
As the observed continuous increase of the CMM mass fraction can be sufficiently well explained by magnetic levitation of gas, it is not surprising to find even colder (denser) material at higher latitudes. This 
can be seen by the unexpectedly large mass fraction of star-forming gas above heights of 300\,pc. Fascinatingly, there is no such gas observed in the hydrodynamic galaxy, supporting the fact that cloud-cloud 
collisions rather result in the dispersion of dense gas than in its levitation.\\
A different way to quantify the effect of magnetic levitation is to determine a typical height of the phases, which we define as
\beq
H_\mathrm{char,phase}=\left(\frac{1}{\Sigma_\mathrm{phase}}\int{z^2\varrho_\mathrm{phase}dz}\right)^{1/2}.
\eeq
Here $\Sigma_\mathrm{phase}$ is the surface density of the respective phase, $\varrho_\mathrm{phase}$ its mass density and $z$ the vertical position. Results for the CMM and star-forming phases 
are given in Fig.~\ref{figCharHeight}. To cover local variations, we show $H_\mathrm{char}$ at three radial distances, namely at $R=\left\{2,5,8\right\}\,\mathrm{kpc}$. \\
Independent of galactocentric distance, the star-forming gas (top row in Fig.~\ref{figCharHeight}) shows larger characteristic heights for the strongest magnetised disc compared with the hydrodynamic 
control disc. At the same time, the height of this phase shows an increasing trend, despite the large fluctuations. This phase reaches heights up to $H_\mathrm{char,hydro}\sim150\,\mathrm{pc}$ without 
magnetic fields, but $H_\mathrm{char,MHD}\sim420\,\mathrm{pc}$ in case of the disc with $\beta=0.25$. The disc with $\beta=1$ shows no pronounced heights at these distances, because the disc fragments 
more efficient farther out.\\
The trend of a larger characteristic height is much more obvious for the CMM-phase. Here, all discs show significant characteristic heights, with the magnitude increasing with 
increasing galactocentric distance. However, the magnetised discs typically show larger heights reaching up to \mbox{$H_\mathrm{char,\beta=0.25}\sim700\,\mathrm{pc}$}. At the same time, the heights of the 
other discs are far smaller, around $H_\mathrm{char}\lesssim200\,\mathrm{pc}$. We point out the similarity between the hydrodynamic disc and the one with $\beta=1$ for annuli $R\ne8\,\mathrm{kpc}$. 
The hydrodynamic disc has fragmented in this regime, but the cold CMM-phase is confined to close to the disc midplane. In contrast, the magnetised disc has not fragmented completely in this range. 
In the fragmented part $R=8\,\mathrm{kpc}$, these discs show diverging heights, with the magnetised disc pushing material higher. Note the rather smooth increase without strong fluctuations.

\section{Brief comparison with previous work}
There are pronounced differences between our approach and the investigations in previous works. One major 
aspect is that we already start with a quite strong magnetic field in a saturated state. In contrast, the works by 
\citet{Pakmor13}, \citet{Rieder16}, \citet{Butsky17}, \citet{Rieder17} or \citet{Steinwandel19} start with a weak seed magnetic field, which 
gets amplified by a small-scale (turbulent) dynamo until it reaches saturation after 1-3\,Gyr. The turbulence in these 
simulations is primarily maintained by supernova feedback, while we explicitly neglected this major feedback 
component. As a consequence, our results do not show large-scale outflows, because no significant (and 
possibly long-term) vertical component of the magnetic field is generated, which would enable material to leave 
the disc and enter the halo. However, for completeness, we mention that \citet{Butsky17} report on a dominant 
toroidal magnetic field in their isolated discs, which suppresses the formation of outflows.\\
Despite the above mentioned differences, the evolution of the magnetic field appears quite similar in our study. 
As shown in Figs.~\ref{figTotMag} and \ref{figBeta}, the magnetic field strength in the diffuse ISM of our galaxies is 
about $B_\mathrm{diffuse}\sim1-10\,\mu\mathrm{G}$ (values of the plasma-$\beta$ have been converted to a field 
strength taking into account the density and temperature values of our cooling curve). As \citet{Pakmor13} point out, 
the saturation value seems to depend on the strength of the initial seed field, but they report on values of 
$B\sim1-5\,\mu\mathrm{G}$. \citet{Steinwandel19} find similar saturation levels, independent of whether the seed field is primordial or injected by supernovae. 

\section{Summary and conclusions}
\label{secSum}
In this study, we present results from numerical simulations of the evolution of disc galaxies with a varying degree of magnetisation. All discs are initialised to possess a constant 
Toomre stability parameter, as well as a constant ratio of thermal to magnetic pressure. The initial magnetic field, if present, is entirely toroidal and scales with the square-root of 
the density. The pressure ratio, $\beta$, is varied between $\beta=0.25$ and $\beta=\infty$, while the Toomre parameter is set to $Q=2$ in the major part of the disc. Stellar feedback via winds, radiation or 
supernovae is not included.\\
Our focus lies on the time evolution of four defined gas phases. This allows us to analyse the disc evolution on larger (the WNM and CNM phases) and smaller (the CMM and star-forming phases) scales. 
The findings of this study can be summarised as follows:\\ 
 
\begin{itemize}
 \item[i)]	Compared to hydrodynamic galaxies, the timescale for instability is increased by the magnetic field, which results in a delayed disc fragmentation.\\
  \item[ii)]	Magnetic fields change the fragmentation pattern as a consequence of a different kind of instability. While the hydrodynamic disc fragments
 radially into rings, the magnetised galaxies fragment into filamentary structures that extend both in the radial and azimuthal direction. This latter result is due 
 to the global Parker instability being the dominant instability in the (stronger) magnetised discs. The dominance arises because of the shorter timescale of the PI compared to the one for gravitational/Toomre instability.\\
 \item[iii)]	Among the magnetised galaxies, the one with the initially lowest $\beta$ (i.e. strongest magnetic field) fragments first. This is the result of the 
 shortest timescale of the PI. \\
  \item[iv)]	Once the discs have fragmented into filaments, and subsequently into individual clouds at later times, the influence of the magnetic field becomes weaker, but non-negligible, and depends on the gas phase. In the 
 warmer phases (WNM and CNM), the impact of the field in the fragmented disc is only marginal. In contrast, the magnetic field allows for a quicker transition to the 
 star-forming phase. In addition, the resulting mass fraction of this phase is larger for the magnetised disc. Hence, magnetic fields channel material to the smallest scales 
 and support it against shear.\\
 \item[v)]	On large scales, the ratio of thermal to magnetic pressure shows only little variation and approximately preserves its initial value. We find $\beta\ll1$ in almost every 
 gas phase. Only discs with $\beta>1$ show values $\beta\sim1$ at late times, but we expect them to evolve below unity.\\
 \item[vi)]	Differences in the plasma-$\beta$ vanish in the star-forming phase, indicating that the clouds decouple from the galactic environment.\\
 \item[vii)]	Beside the global disc thickening due to additional magnetic pressure, we identify a 'general' levitation 
 of dense and cold (and possibly star-forming) material, which is not associated with stellar feedback. The typical heights 
 depend on the phase, but cold material can be as high as $H\sim700\,\mathrm{pc}$. We note that the typical height of 
 the CMM also increases for the hydrodynamic disc. However, this is due to the more frequent cloud-cloud interactions 
 in this disc. 
 

\end{itemize}

\section*{Acknowledgement}
The authors thank the anonymous referee for their insightful report, which helped to improve the quality of this study.
BK thanks M.-A.~Miville-Desch\^{e}nes for providing the IDL routine of the line integral convolution and discussions related to it. BK, RB and WS thank for funding from the DFG grant BA~3706/15-1. BK and RB further appreciate funding from the DFG 
grant BA~3706/4-1. 
REP is supported by a Discovery grant from NSERC - Canada.
BK acknowledges funding via the Australia-Germany Joint Research Cooperation Scheme (UA-DAAD).
The simulations were run on HLRN-III under project grant hhp00043. The \textsc{flash} code was in part developed by the DOE-supported ASC/Alliance Center for
Astrophysical Thermonuclear Flashes at the University of Chicago.

\begin{appendix}

	\section{The disc fragmentation}
\begin{figure}
		\centering
			\includegraphics[width=0.45\textwidth,angle=-90]{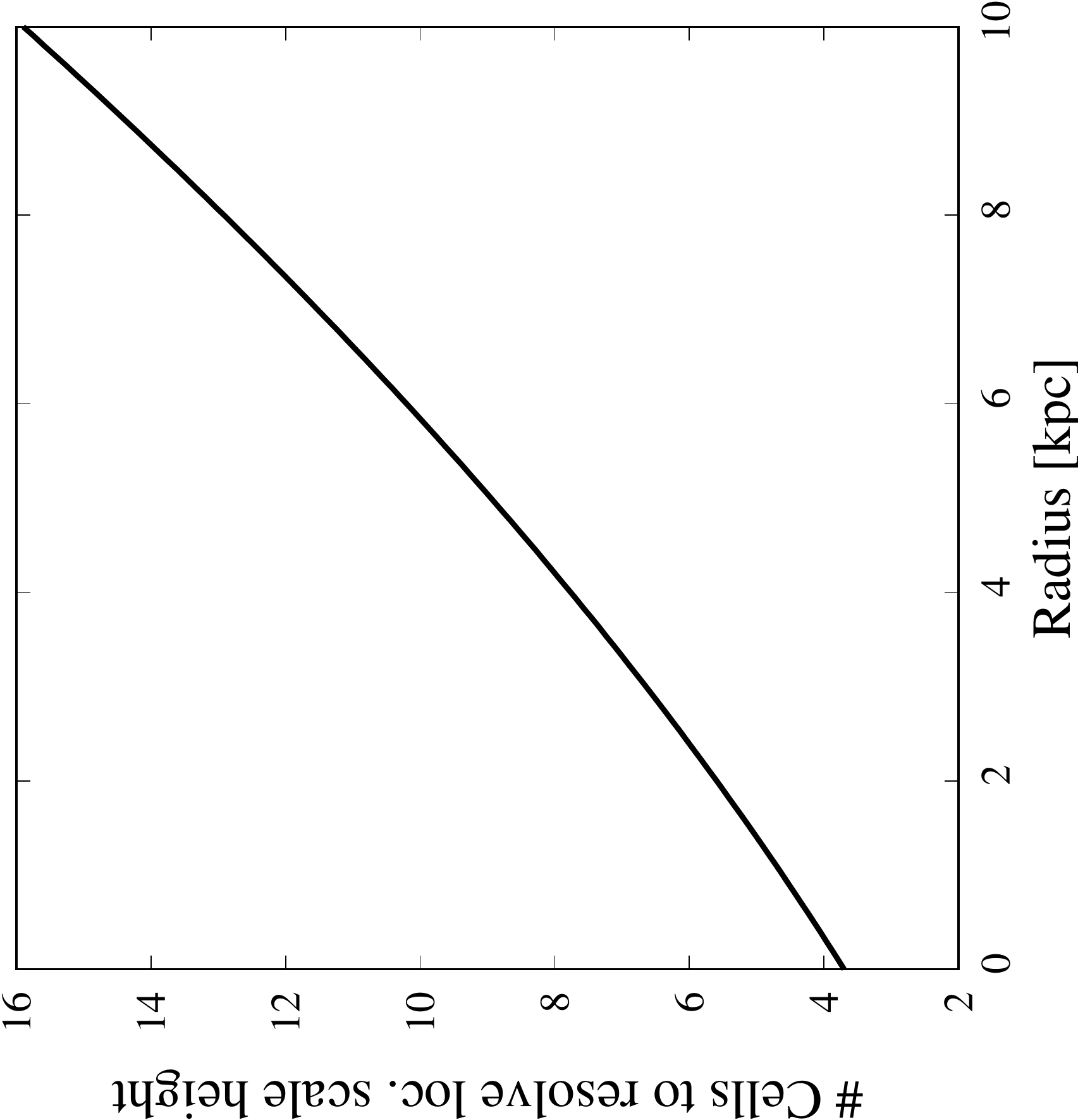}
			\caption{Number of grid cells, which make up the local gas scale height, as a function of radial 
			distance to the galactic center.}
			\label{figAppA}
	\end{figure}

	As has been discussed in \citet{MartinAlvarez18}, a high numerical resolution is necessary to capture the 
	resulting disc dynamics in sufficient detail. We resolve the local Jeans length with at least 32 grid cells, which 
	is sufficient to prevent artificial fragmentation of the gas \citep{Truelove97} as well as to resolve vortical 
	motions within the volume with radius of the Jeans length \citep{Federrath11a}. In addition to our comment in 
	the numerics section, we show in Fig.~\ref{figAppA} by how many grid cells the local gas scale height is resolved. 
	From this figure it is clear that the outer parts of the disc are sufficiently resolved \citep[see also][]{MartinAlvarez18}, 
	while disc radii around $R\sim4-5\,\mathrm{kpc}$ are marginally resolved. The innermost regions of the disc do not 
	resolve the scale height and large-scale fragmentation of the disc might be driven by numerical effects here. However, 
	as is observed in Fig.~\ref{figColDensFO}, fragmentation of the magnetised discs proceeds first in the outer parts, 
	where the scale height is properly resolved. In addition, as stated in \citet{Koertgen18L}, the fragmentation 
	pattern due to the Parker instability is not changed, when turbulent velocity fluctuations are imposed onto the 
	disc rotation to trigger the fragmentation of the disc.
	
	\section{Magnetic field - density scaling}
\begin{figure}
		\centering
			\includegraphics[width=0.385\textwidth,angle=-90]{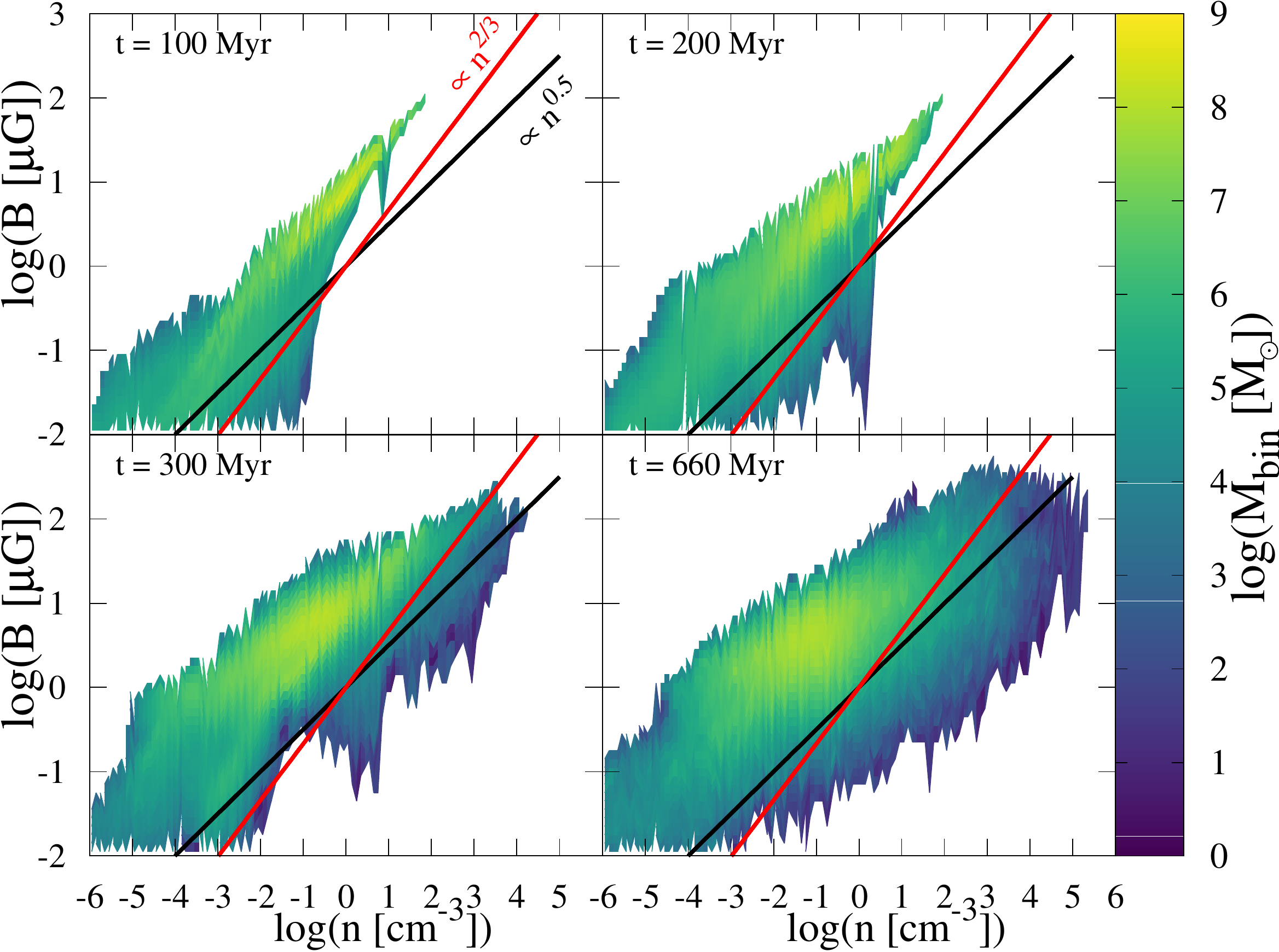}
			\caption{Magnetic field - density relation at different times. Colour coded is the mass per 
			density - field strength bin. Although, the majority of the field strengths follows the initially 
			applied relation $B\propto n^{1/2}$, there appear some deviations in the low-density regime 
			due to adiabatic compression (with a slope of $2/3$). Please note that the spread in 
			field strengths per density bin increases with time. We further emphasise that  
			the bins are shown in colour independent of the number of data points, which populate 
			that bin. }
			\label{figAppB}
	\end{figure}

	A convenient way to study the dynamics of the magnetic field is to analyse its variation across the 
	range of gas densities. A time series of this relation is shown in form of a scatter map in Fig.~\ref{figAppB}. Here, 
	we show in colour the mass within each magnetic field strength - density bin. We remind the reader that our 
	initial scaling was $B\propto n^{1/2}$. This scaling is conserved for the entire duration of the evolution. However, 
	at low densities, deviations of this scaling are identified, which resemble a scaling $B\propto n^{2/3}$. This latter 
	scaling is typical for homologous contraction of gas, which is sometimes also referred to as adiabatic compression. 
	An increase of the field strength with density with a slope $>2/3$ is only marginally observed in the very low-density 
	regime. Such a scaling is indicative of a turbulent dynamo \citep[see also results by][]{Steinwandel19}.

\end{appendix}

\bibliography{astro} 

\bibliographystyle{mn2e}

\end{document}